\documentclass[pra,twocolumn,superscriptaddress,floatfix,showpacs]{revtex4}
\usepackage{amsmath,amsfonts,epsf,dcolumn,natbib,graphicx}
\usepackage{ascii}
\usepackage{xcolor}
\usepackage{soul}
\usepackage{ulem}
\usepackage{mathtools}
\usepackage{verbatim}

\newcommand{\be}{\begin{equation}}
\newcommand{\ee}{\end{equation}}
\newcommand{\beq}{\begin{equation}}
\newcommand{\eeq}{\end{equation}}
\newcommand{\bea}{\begin{eqnarray}}
\newcommand{\eea}{\end{eqnarray}}

\begin{document}
\bibliographystyle{plainnat}
%
\title{
Improved analytical representation of combinations
of Fermi-Dirac integrals for finite-temperature
density functional calculations
}

\author{Valentin V.~Karasiev}
\email{vkarasev@ufl.edu}
\author{Debajit Chakraborty}
\author{S.B.~Trickey}
\affiliation{Quantum Theory Project, 
Departments of Physics and of Chemistry,
University of Florida, Gainesville FL 32611-8435}

\date{06 November 2014}
\begin{abstract}
\noindent 
Smooth, highly accurate analytical representations of 
Fermi-Dirac (FD) integral combinations important in free-energy
density functional calculations are presented.  Specific forms include
those that occur in the local density approximation (LDA),
generalized gradient approximation (GGA), and fourth-order
gradient expansion of the non-interacting free energy as well
as in the LDA and second-order gradient expansion for exchange. By 
construction, 
all the representations and their derivatives of any order 
are continuous on the full domains of their independent variables. 
The same type of technique provides an analytical representation
of the function inverse to the FD integral of order $1/2$.  It plays 
an important role in physical problems
related to the electron gas at finite temperature.
From direct evaluation, the quality of these improved representations 
is shown to be substantially superior to existing ones, many of which
were developed before the era of large-scale computation or early in
the era. 
\end{abstract}


\maketitle

\section{Introduction}
\label{Intro}

Finite-temperature density functional theory (DFT),
whether in orbital-free (OF) or conventional Kohn-Sham (KS) form \cite{Mermin65,Stoitsov88,Dreizler89},
has emerged as a major theoretical and computational tool
for warm dense matter (WDM) studies \cite{Desjarlais02,Horner09,VT84F}. 
In the OFDFT setting,  non-interacting free-energy functionals in 
the local density approximation (LDA) \cite{Feynman..Teller.1949},
gradient-corrected \cite{Perrot.1979,Bartel..Durand.1985}, 
and generalized gradient approximation (GGA) \cite{KST2} forms 
all involve various combinations of Fermi-Dirac (FD) integrals. 
In both OFDFT and KS form, exchange-correlation (XC) functionals  
with explicit temperature (T) 
dependence  are important for 
accurate treatment of the WDM regime \cite{LSDA-PIMC}. 
The finite-temperature LDA exchange energy functional and 
corresponding gradient correction also are 
expressed as combinations of FD integrals
\cite{Horovitz..1974,Geldart.CJP.I,Geldart.CJP.II,Geldart.CJP.III,Geldart.SSC.1994}.
 
Fast, reliable implementation of all these functionals in DFT codes
requires accurate analytical representations of those intrinsic FD 
integral combinations and their derivatives of low order. 
Our experience \cite{KST2}
is that the available representations are not always 
adequate for present-day requirements.  An illustrative difficulty is with
the often-cited representation in  Ref.\ \onlinecite{Perrot.1979}.
Details are below.  A motivating issue is that the representation is
on two sub-domains of the independent variable. 
As a result, the second derivative may behave badly in the
vicinity of the joining of those two pieces.  A second issue is that
some of the available representations were developed before the
era of digital computing or early in it, so that the precision of such
fits (and of the constants in them) is coarse by modern standards.  

Here we present accurate analytical representations for 
FD integral combinations that occur frequently in finite-T DFT
and are important, therefore, to computation.   
The representations are in the form of Pad\'e approximants \cite{Pade}
or Pad\'e approximants modified by additional logarithmic terms.
Some of the parameters for each quantity are constrained to match 
the zero-T and high-T series
expansions for that quantity, with the remaining parameters
determined by fits to accurately evaluated reference data. 
The techniques are similar to those used recently for accurate 
parametrization of the XC energy of the homogeneous electron gas (HEG) 
at finite T \cite{LSDA-PIMC}.
Comparison with some existing 
fits for the same quantities shows that our procedures 
yield  much better accuracy.  

The presentation is organized as follows. Section \ref{FDintegrals}
summarizes FD integral combinations important in finite-T DFT.  
Section \ref{Asymptotics} delineates asymptotic constraints that
are crucial to well-behaved analytical representations of those
combinations.  Section  \ref{Inverse} considers 
the inverse function most relevant to finite-T DFT, 
and its  asymptotic expansions. Section \ref{SecAAR}
describes the analytical representations and accuracy tests 
of them. Parameters
of all the analytical forms are tabulated in the Appendix.
Hartree atomic units are used throughout.

\section{Important Fermi-Dirac integral combinations in finite-T DFT}
\label{FDintegrals}

A key quantity in finite-T OFDFT is the non-interacting 
free-energy density  of the homogeneous electron gas of density $n$.
It is given by the Thomas-Fermi
combination of FD integrals  \cite{{Feynman..Teller.1949}} 
\be
f_{\rm s}^{\rm TF}(n,{\rm T})
=
\frac{\sqrt{2}}{\pi^2 \beta^{5/2}}\Big[-\frac{2}{3}I_{3/2}(\eta)
+\eta I_{1/2}(\eta)\Big]\,,
\label{TF2}
\ee
where $\eta:=\beta\mu$, $\beta := (k_{\mathrm B}{\mathrm T})^{-1}$.  $I_\alpha$ 
is the FD integral \cite{Bartel..Durand.1985}
\bea
I_{\alpha}(\eta ) &:=& \int_0^\infty dx \frac{x^\alpha}{1 + \exp (x - \eta)} %
\; , \;\;\; \alpha > -1\nonumber \\
I_{\alpha -1} (\eta ) &=&\frac{1}{\alpha} \frac{d}{d \eta} I_\alpha (\eta )
\,,
\label{FDintDefn}
\eea
and $\mu$ is the chemical potential defined by the density $n$ as
\be
n=\frac{\sqrt{2}}{\pi^2\beta^{3/2}}I_{1/2}(\beta\mu) \,.
\label{TF1b}
\ee
\lbrack Remark: There are at least two conventional definitions of
FD integrals as well as variations.  The definition used here 
\cite{Bartel..Durand.1985,Blakemore1982}  
is related to the one used by Huang \cite{HuangText}, denoted  
as $f_{3/2}(w) = (2/{\sqrt \pi}) I_{1/2}(\ln w) $. \rbrack

It is convenient to define the  reduced temperature
\be
t=\mathrm{T}/\mathrm{T}_{\rm F}=\frac{2}{\beta [3\pi^2n]^{2/3}}\, 
\label{tred}
\ee
in terms of which Eq. (\ref{TF1b}) becomes 
\be
I_{1/2}(\beta\mu) = \frac{2}{3t^{3/2}}  \; .
\label{TF1c}
\ee
Since $I_{1/2}(\eta ) $ is a strictly increasing function
of $\eta = \beta \mu$, it follows that all functions of $\eta$
are functions of the reduced temperature $t$.  
Sometimes a related variable
\be
y(\eta) := I_{1/2}(\eta) \equiv \frac{2}{3t^{3/2}} \,.
\label{y} 
\ee
is used instead of $t$.

Returning to 
Eq.\ (\ref{TF2}), we have 
\be
f_{\mathrm{s}}^{\mathrm{TF}}(n,\mathrm{T})=\frac{n}{\beta}f(\eta)
\equiv \tau_0^{\mathrm{TF}}(n)\kappa(\eta)
\,,
\label{TF}
\ee
where $\tau_0^{\rm TF}(n)=(3/10)(3\pi^2)^{2/3}n^{5/3}$
and 
\bea
f(\eta) & := & \frac{1}{I_{1/2}(\eta)} \Big[-\frac{2}{3}I_{3/2}(\eta)
+\eta I_{1/2}(\eta)\Big] \,,
\label{f}
\\
\kappa(\eta) & := &\frac{5\times 2^{2/3}}{3^{5/3}} 
\frac{1}{I_{1/2}^{5/3}(\eta)} 
\Big[-\frac{2}{3}I_{3/2}(\eta)
+\eta I_{1/2}(\eta)\Big] \,.
\label{kappa}
\eea
The functions $f$ and $\kappa$ have a simple relation
\be
\kappa (\eta) =\frac{5}{3}t\,f(\eta)  \; .
\label{kappa-f} 
\ee

These two comparatively simple functions, together with the inverse 
problem for Eq.\ (\ref{y}) (to find $\eta(y)$), illustrate the 
issue addressed here.  Computational performance demands accurate
analytical representations of such quantities.  Achieving such 
representations for $\kappa$ or $f$ as a function of $\eta$ or $y (\eta)$ 
requires development of suitably constrained analytical forms and 
fitting of them to accurately calculated reference values. 
An example is the work of Perrot, who used the variable $y$ 
and provided an analytical fit for 
$f(y)$ on two $y$ intervals \cite{Perrot.1979}.   Use of 
two intervals is an issue noted already and considered further below.

Another important category of FD integral combinations arises from
density-gradient contributions.   
The finite-T generalized gradient approximation (GGA) \cite{KST2} for
the non-interacting free-energy  uses T-dependent variables defined 
by analysis of the second-order gradient expansion (SGE) 
of the non-interacting free energy density of the weakly inhomogeneous
electron gas \cite{Perrot.1979,Geldart.Sommer.1985}.  The gradient
term in the SGE has the coefficient 
\be
f_{\mathrm{s}}^{(2)}(n,\nabla n,\mathrm{T})=\tau_0^{\mathrm{TF}}(n)\frac{5}{27} s^2\widetilde B
\,,
\label{2nd}
\ee
where $s=|\nabla n|/2(3\pi^2)^{1/3}n^{4/3}$,
along with the definition 
\be
{\widetilde B}(\eta) := - 3\frac{I_{1/2}(\eta) I_{-3/2}(\eta)}%
{I_{-1/2}^2(\eta)}\,.
\label{B}
\ee
High-quality representation of ${\widetilde B}$ thus is required.
\lbrack Remark: the quantity $\widetilde B$ defined here is $\tilde h$ in 
the notation of Ref.\ \cite{KST2}. \rbrack

The fourth-order term in the gradient expansion for the non-interacting 
free energy density derived in Refs.\ 
\cite{Geldart.Sommer.1985,Geldart.Sommer.1985B,Bartel..Durand.1985} 
takes the form
\be
f_{\mathrm{s}}^{(4)}(n,\nabla n, \nabla^2 n,\mathrm{T})
=\tau_0^{\mathrm{TF}}(n)
\Big[
\frac{8}{81}p^2 \widetilde C - \frac{1}{9}s^2 p \widetilde D + 
\frac{8}{243}s^4 \widetilde E
\Big]
\,,
\label{4th}
\ee
where $p:=\nabla^2n/4(3\pi^2)^{2/3}n^{5/3}$.  The ingredient 
combinations of FD integrals are 
\be
\widetilde C(\eta) :=\frac{5\times 3^{11/3}}{2^{11/3}} I_{1/2}^{5/3}(\eta)
\Big[
\frac{1}{9}\frac{I_{-3/2}^2(\eta)}{I_{-1/2}^3(\eta)}
-\frac{1}{5}\frac{I_{-5/2}(\eta)}{I_{-1/2}^2(\eta)}
\Big]
\,,
\label{C}
\ee
\bea
\widetilde D(\eta) :=&&\frac{5\times 2^{1/3}}{3^{1/3}} I_{1/2}^{8/3}(\eta)
\Big[
-3\frac{I_{-7/2}(\eta)}{I_{-1/2}^3(\eta)}
\nonumber\\
&& + \frac{33}{10}\frac{I_{-3/2}(\eta)I_{-5/2}(\eta)}{I_{-1/2}^4(\eta)}
-\frac{I_{-3/2}^3(\eta)}{I_{-1/2}^5(\eta)}
\Big]
\,,
\label{D}
\eea
and
\bea
\widetilde E(\eta) :=&&\frac{5\times 3^{14/3}}{2^{2/3}} I_{1/2}^{11/3}(\eta)
\Big[
-\frac{7}{96}\frac{I_{-9/2}(\eta)}{I_{-1/2}^4(\eta)}
\nonumber\\
&& -\frac{1}{15}\frac{I_{-3/2}^{2}(\eta)I_{-5/2}(\eta)}{I_{-1/2}^6(\eta)}
+\frac{1}{72}\frac{I_{-3/2}^4(\eta)}{I_{-1/2}^7(\eta)}
\nonumber\\
&&+ \frac{1}{12}\frac{I_{-3/2}(\eta)I_{-7/2}(\eta)}{I_{-1/2}^5(\eta)}
+\frac{1}{32}\frac{I_{-5/2}^2(\eta)}{I_{-1/2}^5(\eta)}
\Big]
\,.
\label{E}
\eea

Exchange and correlation also are expressed in terms of FD integral
combinations. For the weakly inhomogeneous
electron gas at finite T,
the LDA exchange (X) free-energy density \cite{Perrot.1979,Horovitz..1974}
is 
\be
f_{\rm x}^{\mathrm{LDA}}(n,{\rm T})=
-\frac{1}{2\pi^3\beta^2}\int_{-\infty}^{\eta}[I_{-1/2}(\eta)]^2 d\eta
\, ,
\label{fxLDA}
\ee
or 
\be
f_{\rm x}^{\mathrm{LDA}}(n,{\rm T})= \widetilde A_{\mathrm{x}}(n,{\mathrm T}) %
e_{\rm x}^{\rm LDA}(n)
\,,
\label{fxLDA2}
\ee
Here $e_{\rm x}^{\rm LDA}(n)=-\frac{3}{4}\Big(\frac{3}{\pi}\Big)^{1/3}n^{4/3}$
is the zero-T LDA X energy density (evaluated, of course, with the
finite-T density) and the relevant FD integral combination is 
\be
\widetilde A_{\mathrm{x}}(\eta):=
\frac{f_{\rm x}^{\mathrm{LDA}}(n,{\rm T})}{e_{\rm x}^{\rm LDA}(n)}
=\frac{2^{1/3}}{3^{4/3}}\frac{\int_{-\infty}^{\eta}[I_{-1/2}(\eta)]^2 d\eta}{I_{1/2}^{4/3}(\eta)}
\,.
\label{Ax}
\ee
The corresponding expressions for 
the second-order gradient correction to the LDA exchange free-energy  
\cite{Geldart.CJP.I,Geldart.CJP.II,Geldart.CJP.III,Geldart.SSC.1994} 
(see also \cite{Geldart.TCC.1996}) are 
\be
f_{\rm x}^{(2)}(n,\nabla n,{\rm T})=
e_{\rm x}^{\rm LDA}(n)
\frac{8}{81}s^2\widetilde B_{\mathrm{x}} \, ,
\label{fxSGA2}
\ee
with 
\be
\widetilde B_{\mathrm{x}}(\eta) :=
  \frac{3^{4/3}}{2^{4/3}} I_{1/2}^{4/3}(\eta)
\Big[\Big(\frac{I_{-1/2}'(\eta)}{I_{-1/2}(\eta)}\Big)^2
-3\frac{I_{-1/2}''(\eta)}{I_{-1/2}(\eta)}\Big]\, .
\label{Bx}
\ee
Primes denote derivatives with respect to the argument.

Practical implementation of all the combinations of FD integrals
discussed above and implementation of density functionals
with explicit T-dependence based on variables related to 
these combinations (for example, the T-dependent variables
related to $\widetilde B$ in the GGA non-interacting free-energy \cite{KST2}), 
require accurate analytical
representations for the quantities in Eqs.\ (\ref{f}) or (\ref{kappa})), 
(\ref{B}), (\ref{C})-(\ref{E}), (\ref{Ax}), and (\ref{Bx}).  

For ease of use of the results, 
Table \ref{tab:table0} provides a cross-reference of all these quantities  
to both the analytical
forms and the tables of coefficients provided in the fits we present here. 
Note that the quantity $\eta_{1/2}$ in the last line of Table \ref{tab:table0}  
is the function inverse to $I_{1/2}(\eta)$.  Detail about it is 
in Sec.\ \ref{Inverse}.

\begin{table}
\caption{\label{tab:table0}
Cross-reference of equation numbers for quantities considered in 
the present work
and the corresponding analytical
forms for fits and for Tables of coefficients for those fits.
}
\begin{ruledtabular}
\begin{tabular}{cccc}
Quantity & Definition & Analytical form & Coefficient values \\
\hline
$\widetilde B$              & Eq. (\ref{B}) & Eq. (\ref{Bfit}) & Table \ref{tab:table2} \\
$\widetilde C$              & Eq. (\ref{C}) & Eq. (\ref{CDEfit}) & Table \ref{tab:table3} \\
$\widetilde D$              & Eq. (\ref{D}) & Eq. (\ref{CDEfit}) & Table \ref{tab:table4} \\
$\widetilde E$              & Eq. (\ref{E}) & Eq. (\ref{CDEfit}) & Table \ref{tab:table5} \\
$\widetilde A_{\mathrm{x}}$ & Eq. (\ref{Ax}) & Eq. (\ref{Axfit}) & Table \ref{tab:table6} \\
$\widetilde B_{\mathrm{x}}$ & Eq. (\ref{Bx}) & Eq. (\ref{Bxfit}) & Table \ref{tab:table7} \\
$\eta_{1/2}$                & Eq. (\ref{etadefn}) & Eq. (\ref{etafit}) & Table \ref{tab:table8} \\
\end{tabular}
\end{ruledtabular}
\end{table}

\section{Asymptotic behaviors as constraints}
\label{Asymptotics}

Knowledge of the asymptotic forms for the combinations just discussed 
is important for constraining analytical representations of them.  
Without such constraints, representations fitted over a finite 
independent-variable domain become uncontrolled approximations 
outside that domain.  There are two relevant limits.

In the non-degenerate limit ($\eta \ll -1 $), which corresponds to 
$y\rightarrow 0$, the functions given in  Eqs.\ (\ref{f}), 
(\ref{kappa}), (\ref{B}), (\ref{C})-(\ref{E}), (\ref{Ax}) and (\ref{Bx})
have the following asymptotic expansions in terms of the variable $y$:
\bea
f(y)&=&-1+\ln(\frac{2}{\sqrt{\pi}}y) + O(y)\,,
\nonumber\\
\kappa(y)&=&\frac{5\times 2^{2/3}}{3^{5/3}}y^{-2/3}\Big(-1+ %
\ln(\frac{2}{\sqrt{\pi}}y)\Big)+O(y^{1/3}) \,,
\nonumber\\
\widetilde B(y)&=&3-\frac{3y}{\sqrt{2\pi}}+O(y^2) \,,
\nonumber\\
\widetilde C(y)&=&\Big(\frac{3}{2}\Big)^{5/3}y^{2/3}+O(y^{5/3})\,,
\nonumber\\
\widetilde D(y)&=&\Big(\frac{2}{3}\Big)^{1/3}y^{2/3}+O(y^{5/3})\,,
\nonumber\\
\widetilde E(y)&=&\frac{3^{8/3}}{2^{25/6}}\frac{y^{5/3}}{\sqrt{\pi}}+O(y^{8/3})\,,
\nonumber\\
\widetilde A_{\mathrm{x}}(y)&=&  \frac{2^{4/3}}{3^{4/3}}y^{2/3}+O(y^{5/3}) \,,
\nonumber\\
\widetilde B_{\mathrm{x}}(y)&=&  -\frac{3^{4/3}}{2^{1/3}}y^{4/3}+O(y^{7/3}) \,,
\label{small-y}
\eea

Bearing in mind that $y=y(\eta)$, the foregoing relationships are derived 
as follows. 
The series expansion for $I_{\alpha}(\eta)$ ($\alpha>-1$, $\eta\le 0$)
(see \cite{Cox.Giuli.1968,Cloutman.1989} for details) is 
\be
I_{\alpha}(\eta)=\Gamma(\alpha+1)e^{\eta}\sum_{k=0}^{\infty}
(-1)^{k}\frac{e^{k\eta}}{(k+1)^{\alpha+1}}
\,.
\label{small-y2}
\ee
In the non-degenerate limit, the leading terms of Eq.\ (\ref{small-y2})
for $\alpha=1/2$ are
\be
y \equiv I_{1/2}(\eta) \approx \frac{\sqrt{\pi}}{2}e^{\eta}
\Big(1-\frac{e^{\eta}}{2\sqrt{2}}\Big)
\,.
\label{small-y3}
\ee
After series expansion of the negative solution to this quadratic in
$e^\eta$, one has the inversion
\be
\eta(y)=\ln(\frac{2}{\sqrt{\pi}}y) + \frac{y}{\sqrt{2\pi}}+O(y^2)
\,.
\label{small-y4}
\ee

Next, consider the first two leading terms in the 
non-degenerate limit
for FD integrals with indices $\alpha=3/2,-1/2$  
from Eq.\ (\ref{small-y2}).  For  those with 
indices $\alpha=-3/2,-5/2,-7/2$, and $-9/2$, we first use the
recursion relation Eq.\ (\ref{FDintDefn}), then substitute the 
result from Eq.\ (\ref{small-y3}) to obtain
\bea
I_{3/2}(\eta) &\approx& 
\frac{3\sqrt{\pi}}{4} e^{\eta}(1-\frac{e^{\eta}}{4\sqrt{2}})
\,,
\nonumber\\
I_{-1/2}(\eta) &\approx& 
\sqrt{\pi}e^{\eta}(1-\frac{e^{\eta}}{\sqrt{2}})
\,,
\nonumber\\
I_{-3/2}(\eta) &\approx& 
-2\sqrt{\pi}e^{\eta}(1-\sqrt{2}e^{\eta})
\,.
\label{small-y5}
\eea
\lbrack Remark: Eq.\ (\ref{small-y2}) also provides the leading terms 
correctly for $\eta<<-1$ independent of $\alpha$).\rbrack \    
Elimination of the $\eta$ variable in these leading terms
via Eq.\ (\ref{small-y4}),
substitution into equations for the quantities of interest, 
e.g.\ $f$, $\kappa$, $\widetilde B$
etc. defined by Eqs.\ (\ref{f})-(\ref{kappa}), (\ref{B}),
(\ref{C})-(\ref{E}), (\ref{Ax})-(\ref{Bx}),
and subsequent small-$y$ series expansion up through the 
first $y$-dependent term (if necessary) yields 
Eqs.\ (\ref{small-y}).
\lbrack Remark: It is important to note that  only the expansion 
for $\widetilde B$ requires  use of the two leading terms in the 
FD series and subsequent small-$y$ series expansion, but even in that
specific case, the second term in Eq.\ (\ref{small-y4})
does not contribute to the final result, thus can be dropped in 
the process of elimination of the $\eta$ variable.
All the other expansions in Eq.\ (\ref{small-y}) can be obtained 
with just the first term in Eqs.\ (\ref{small-y3})-(\ref{small-y5}).\rbrack
 
Conversely, in the degenerate limit $y\rightarrow \infty$ ($\eta \gg 1$),
the leading terms in the asymptotic expansions of the FD 
integral combinations under consideration are 
\bea
f(y)&=&\frac{3}{5}\Big(\frac{3}{2}\Big)^{2/3}y^{2/3}+O(y^{-2/3}) \,,
\nonumber\\
\kappa(y)&=&1-\frac{5}{2^{2/3}3^{7/3}}\pi^2y^{-4/3}+O(y^{-8/3})\,,
\nonumber\\
\widetilde B(y)&=&1+\frac{2^{4/3}}{3^{7/3}}\pi^2 y^{-4/3}+O(y^{-8/3})\,,
\nonumber\\
\widetilde C(y)&=&1+\frac{17}{2^{5/3}3^{7/3}}\pi^2y^{-4/3}+O(y^{-8/3})\,,
\nonumber\\
\widetilde D(y)&=&1+\frac{413}{2^{2/3}3^{16/3}}\pi^2y^{-4/3}+O(y^{-8/3})\,,
\nonumber\\
\widetilde E(y)&=&1+\frac{47}{2^{5/3}3^{7/3}}\pi^2y^{-4/3}+O(y^{-8/3})\,,
\nonumber\\
\widetilde A_{\mathrm{x}}(y)&=& 1-\frac{2^{4/3}}{3^{10/3}}\pi^2y^{-4/3}
\Big(\ln(y)+const.\Big) +O(y^{-8/3}) \,,
\nonumber\\
\widetilde B_{\mathrm{x}}(y)&=& 1+\frac{1}{2^{2/3}3^{1/3}}\pi^2y^{-4/3}+O(y^{-8/3})\,,
\label{large-y}
\eea

The derivation of these expressions is essentially the same as for
the opposite limit.  The  asymptotic series for $I_{\alpha}(\eta)$, 
now for $\eta>0$, is (see Ref.\ \cite{GaroniFrankelGlasser2001} Eq.\ (3.10),
but note that they use the Dingle \cite{Dingle.1957} normalization for 
the FD integrals)
\bea
I_\alpha(\eta)& \sim & \Gamma(\alpha + 1) \biggr\lbrace %
\cos({\alpha \pi}) I_\alpha(-\eta)   \nonumber \\
&& \left. + \sum_{\nu = 0}^\infty \frac{2\tau_{2\nu}}{\Gamma(\alpha+2 - 2\nu)} %
\eta^{\alpha+1-2\nu}\right\rbrace \nonumber \\
\tau_n &:=&\sum_{m=1}^\infty \frac{(-1)^{m+1}}{m^n} = %
\lbrack 1 - 2^{(1-n)}\rbrack \zeta(n)
\label{GFG3.10} 
\eea
with $\zeta(n)$ the Riemann zeta function.  (Also see Ref.\ 
\cite{Dingle.1957}) and  Eq.\ (5) in 
Ref.\ \cite{Cody.Thacher.1967}.)  Thus, the equation 
analogous to Eq.\ (\ref{small-y3}) in the degenerate limit is
\be
I_{1/2}(\eta) \approx \frac{2}{3}\eta^{3/2}
+\frac{\pi^2}{12}\eta^{-1/2}
\,.
\label{large-y2}
\ee
The second term may be dropped, which gives the inversion
\be
\eta(y)=\Big(\frac{3}{2}\Big)^{2/3} y^{2/3}
\,.
\label{large-y3}
\ee
Correspondingly, the leading terms in the degenerate limit of other 
relevant FD integrals are 
\bea
I_{3/2}(\eta) &\approx & \frac{2}{5}\eta^{5/2} 
+  \frac{\pi^2}{4}\eta^{1/2}
\,,
\nonumber\\
I_{-1/2}(\eta) &\approx& 2\eta^{1/2} 
-  \frac{\pi^2}{12}\eta^{-3/2}
\,,
\nonumber\\
I_{-3/2}(\eta) &\approx& -2\eta^{-1/2} 
-  \frac{\pi^2}{4}\eta^{-5/2}
\,.
\label{large-y4}
\eea
Again, elimination of the $\eta$ variable in these leading terms, 
substitution into 
the equations for the quantities of interest, and one more series expansion
(if needed) in the large-$y$ limit up through the first $y$-dependent term 
yields the large-$y$ asymptotic expressions of Eqs.\ (\ref{large-y}).

Some of the results presented in Eqs.\ (\ref{small-y}) and (\ref{large-y})
can be found in Refs.\ \cite{Perrot.1979,Geldart.Sommer.1985B}.
Comparison between accurately evaluated reference data 
(see Sec. \ref{subsecPresW}) and values 
obtained with terms in Eqs.\ (\ref{small-y}) and (\ref{large-y})
at extremely small-$y$ and large-$y$ also was used for verification 
of the expansion coefficients in Eqs.\ (\ref{small-y}) and (\ref{large-y}).

\section{Inverse functions}
\label{Inverse}

We have already mentioned the importance of functions inverse to 
Fermi-Dirac integrals.  These arise, for example, 
in problems related to the description of the electron gas at finite-T.
Perhaps the most familiar example is the solution of Eq.\ (\ref{TF1b})
or, equivalently, of Eq.\ (\ref{y}).  Either requires  
the inverse to the function $I_{1/2}(\eta)$ to obtain the 
dimensionless Fermi energy $\eta=\beta\mu$.  For the sake of
generality,  we define
the inverse function for an FD integral of order $\alpha$ as
\be
\eta_{\alpha}(I_{\alpha}(\eta)) :=\eta
\,.
\label{etadefn}
\ee
The only one of interest here is  
\be
\eta_{1/2}(y)=\eta 
\,.
\label{eta}
\ee
Various analytical fits (as a function of 
the $y$ variable) to the 
solution of this specific inversion problem have been proposed 
\cite{Nilsson.1973,Blakemore1982}.
The leading terms in the series expansion for $\eta_{1/2}$
in the non-degenerate and degenerate limits are given by Eqs.\ (\ref{small-y4})
and (\ref{large-y3}) respectively, 
\bea
\eta_{1/2}(y) &=& \ln(\frac{2}{\sqrt{\pi}}y) + O(y)
\,; y\rightarrow 0
\,,
\nonumber\\
\eta_{1/2}(y) &=& \Big(\frac{3}{2}\Big)^{2/3} y^{2/3}+O(y^{-2/3}) 
\,; y\rightarrow \infty
\,.
\label{small-large-y-inv}
\eea
Analogous terms in the small-$y$ and large-$y$ expansions for 
other inverse functions $\eta_{\alpha}(y)$ 
could be obtained by inversion of Eqs.\ (\ref{small-y5}) and
(\ref{large-y4}).  We have not needed such inverse
functions, so do not consider them here.

\section{Smooth analytical representations}
\label{SecAAR}
\subsection{Context}
\label{subsecPriorW}
In Ref.\ \cite{Perrot.1979} Perrot provided analytical representations
for both $f$ and $\widetilde B$.   Ref.\ \cite{Geldart.Sommer.1985}
gave fits for $\widetilde B$, \ldots,  $\widetilde E$.
Both works used  least squares fits (LSF) of Chebyshev 
polynomials to tabulations of calculated FD integrals. Such fits 
rely upon reference data on a finite interval of the 
variable $\eta\in [\eta_{\mathrm{min}},\eta_{\mathrm{max}}]$.  This is 
equivalent to mapping $y\in [y_{\mathrm{min}},y_{\mathrm{max}}]$  
onto the polynomial argument on the interval $[-1,1]$.
In Ref.\ \cite{Perrot.1979}, the data are tabulated on 
the interval $\eta \in [-10,20]$.  The corresponding $y$ variable 
interval was divided into  $[4\cdot 10^{-5},y_0]$ and $[y_0,59.8]$, where
$y_0=3\pi/4\sqrt{2}$.  At $y_0$, the function and its 
first and second derivatives were required to be continuous. 
The advantage of using two intervals is that the small-$y$ and large-$y$ 
asymptotic behaviors can be incorporated easily in the fits, thereby 
making them applicable for the entire range $y\in [0,\infty[$. The 
disadvantage is that the fits turn out not to be smooth near $y_0$ despite 
the enforcement of continuity.  Numerical illustrations of the issue
in the case of the fit to 
$\widetilde B$ from Ref.\ \cite{Perrot.1979}
are provided in the next sub-Section.  Note that the  
fit to $f$ is smooth through the second derivative.

For the second- and fourth-order gradient corrections to the non-interacting 
free energy, Geldart and Sommer \cite{Geldart.Sommer.1985} also used 
Chebyshev polynomials, but fitted to 
tabulated data on the entire interval $y\in [0.02,25]$.  They thereby avoided 
the problem of connection at an intermediate point. The drawback 
is that there is no straightforward means to incorporate the 
asymptotic behaviors beyond the ends of the interval. 

Ref.\ \cite{PDW84} provided a fit to the LDA exchange Eq.\ (\ref{fxLDA})
in the form of a Pad\'e approximant of order [4,4] in terms of 
the variable $t$. The accuracy of that fit is examined below.

A simple fit to the inverse $\eta_{1/2}$ function
is given by Eq.\ (8) of Ref.\ \cite{Nilsson.1973} 
(see also Eq.\ (38) in Ref.\ \cite{Blakemore1982} and Table \ref{tab:table1}
below for the fit accuracy). 
Ref.\ \cite{Antia.1993} provides a rational function approximation
for inverse functions of selected FD integrals. 
Though those fits are accurate, the approximations again are 
on two separate domains of the corresponding independent 
variable. That immediately brings into play the possibility of
continuity problems with functional derivatives. Also, the most accurate
approximation provided in Ref.\ \cite{Antia.1993}
has a significantly larger number of parameters than the 
analytical representation given in the next subsection. 

\subsection{Present work - analytical representations and fitting}
\label{subsecPresW}
In contrast with prior practice,  we used both a standard LSF 
and augmented non-linear fits
(discussed below) to data we calculated on a large interval, 
$\eta \in [-11,100]$.  That 
corresponds to $y\in [1.48\cdot 10^{-5},667]$ and $t\in [0.01,1266]$. 
The tabulation was on a uniform $\eta$ mesh with increment 
$\Delta \eta = 0.025$.    Excluding two boundary points from each end 
gives a total of 4436 mesh points. 
FD integrals of order $\alpha=3/2$, $1/2$, $-1/2$ and their first and 
second derivatives were calculated by the quadrature methodology presented 
in Ref.\  \cite{FD.Aparacio.1998}.  The version that uses 80 evaluations 
of the integrand and guarantees about 15 decimal digits accuracy was employed. 

For $\alpha=-3/2$, $-5/2$, $-7/2$, and $-9/2$, the FD integrals were calculated 
using the recursion relation Eq.\ (\ref{FDintDefn}) with commuted
differentiation and integration,
\be
I_{\alpha-k}(\eta ) = \frac{\Gamma(\alpha)}{\Gamma(\alpha-k+2)} \int_0^\infty dx 
\Big(\frac{d}{d \eta}\Big)^k 
\frac{x^\alpha}{1 + \exp (x - \eta)} %
\,
\label{FDint2}
\ee
with $\alpha=-1/2$ and $k=1,2,3,4$. 
The derivatives were done analytically and the integration
by the same quadrature method as for the other indices.

The FD integral in the numerator of 
$\widetilde A_{\mathrm{x}}(\eta)$, Eq.\ (\ref{Ax}), was 
separated into two parts, 
\be
\widetilde A_{\mathrm{x}}(\eta)=
\frac{2^{1/3}}{3^{4/3}}
\Big[
\frac{\int_{-\infty}^{\eta_1}[I_{-1/2}(\eta)]^2 d\eta
+
\int_{\eta_1}^{\eta}[I_{-1/2}(\eta)]^2 d\eta}
{I_{1/2}^{4/3}(\eta)}
\Big]
\,,
\label{AxIntEval}
\ee
where $\eta_1$ is the first mesh point ($\eta_1=-11$ 
which corresponds to $y_1\equiv y(\eta_1)=1.48\cdot 10^{-5}$).
The first integral in this expression was calculated using 
the small-$y$ series expansion for $\widetilde A_{\mathrm{x}}$
(see Eq.\ (\ref{small-y})),
$\int_{-\infty}^{\eta_1}[I_{-1/2}(\eta)]^2 d\eta=
3^{4/3}2^{-1/3}\widetilde A_{\mathrm{x}}(y(\eta_1))I_{1/2}^{4/3}(\eta_1)$.
The second term was evaluated numerically on the uniform $\eta$ mesh 
described above using Simpson's rule with additional evaluation
of the integrand at intermediate points $\eta_i+\Delta\eta/2$.
Derivatives of all FD integral combinations with respect
to the variable $y$ were done numerically.

Regarding the analytical forms to represent the FD integral combinations
of interest, 
greater flexibility is required 
than is provided by Chebyshev polynomials
if the leading
asymptotic terms for both small- and large-$y$ limits are to 
be incorporated.  Moreover, the resulting representations 
should be intrinsically continuous (including all derivatives)
on the entire range of $y\in [0,\infty[$.  These considerations
led to the adoption of Pad\'e approximants \cite{Pade} for
the basic representations. As will become apparent, augmentation
is required in some cases.  The method can be sketched as follows: 
(i) select appropriate variables for two polynomials in the form $y^{m/3}$,
where $m=1,2,3$ or 4;
(ii)  constrain appropriate coefficients to match small-$y$ and large-$y$ series
expansions; (iii) fit the remaining coefficients.  

Because the fits to $\kappa$,
$\widetilde A_{\mathrm{x}}$ and $\eta_{1/2}$ require use of modified approximants 
which incorporate logarithmic terms, it is simplest to begin with
$\widetilde B$.  
The leading terms of its asymptotic expansions,
Eqs.\ (\ref{small-y}) and (\ref{large-y}), dictate the choice
of fitting function to be
\be
\widetilde B(y) = \frac{\sum_{i=0}^{8}a_iy^i}
{1+\sum_{i=1}^{12}b_iu^i}
\,,
\label{Bfit}
\ee
where $u=y^{2/3}$. This is a Pad\'e approximant 
of order $[24,24]$ with respect to the variable $y^{1/3}$.
Correct powers of the first few terms in the small-$y$ and large-$y$
asymptotic expansions are assured by setting 
$a_7= b_1= b_2= b_4= b_{11}=0$.  Doing so eliminates terms in
$y^{2/3}$, $y^{4/3}$, $y^{5/3}$, $y^{7/3}$, $y^{8/3}$, and in 
$y^{-2/3}$, $y^{-1}$, $y^{-5/3}$, $y^{-7/3}$ 
from the small- and large-$y$ asymptotic expansions respectively.
Agreement with the correct coefficients for the 
leading small-$y$ expansion terms requires 
$a_0=3$ and $a_1=-3/\sqrt{2\pi}$.  Similarly, the
large-$y$ asymptotic expansion coefficients are matched by setting 
$b_{12}=a_8$ and $b_{10}=-2^{4/3}3^{-7/3}\pi^2b_{12}$. The remaining
coefficients could, in principle, be set by LSF but that turns out
not to be best.  See discussion below.   

Turning next to $\widetilde C$, $\widetilde D$, and $\widetilde E$,
because their small-$y$ expansions have terms proportional to powers 
of $y^{1/3}$ (see Eqs.\ (\ref{small-y}) and (\ref{large-y})), 
the fitting functions are of the  form
\be
R(y) = \frac{a_{2.5}u^{5/2}+\sum_{i=1}^{12}a_iu^i}
{1+\sum_{i=1}^{6}b_iv^i}
\,,
\label{CDEfit}
\ee
where $v=y^{4/3}$.  The term $\propto u^{5/2}$ is added to match  
the $y^{5/3}$ term present in the small-$y$ expansion. 
Eq.\ (\ref{CDEfit}) has the form of a [24,24] Pad\'e approximant 
 with respect to the variable $y^{1/3}$. Setting 
$a_{11}=0$ in Eq.\ (\ref{CDEfit}) for 
$\widetilde C$ and $\widetilde D$
eliminates the $y^{-2/3}$ term in the large-$y$
asymptotic expansion. We found it beneficial to keep terms in 
$y^{4/3}$ and $y^{-2}$ in the small-$y$ and 
large-$y$ expansions correspondingly, in that doing so reduces fitting 
errors considerably.  
Setting $a_1= a_2 = a_{11}=0$ for $\widetilde E$
eliminates terms proportional to $y^{2/3}$ and $y^{-2/3}$
in the asymptotic expansions. Again, keeping $y^{2}$ and $y^{-2}$ terms in
the small- and large-$y$ expansions
led to reduced fitting errors.
Additional constraints on the parameters in Eq.\ (\ref{CDEfit})
arise from fixing the coefficients of the leading terms in 
small-$y$ and large-$y$ expansions Eqs.\ (\ref{small-y}) and 
(\ref{large-y})) (see corresponding Tables in the Appendix). 

Fitting of $\widetilde A_{\mathrm{x}}(y)$ 
requires incorporation of a logarithmic term to satisfy
the large-$y$ asymptotic expansion (see Eq.\ (\ref{large-y})).
The modified Pad\'e approximant of order [16,16] (with respect to
the variable $y^{1/3}$) is similar to Eq.\ (\ref{CDEfit}) (with the additional 
$\log$ term in the numerator), 
\be
\widetilde A_{\mathrm{x}}(y) = \frac{a_{ln}y^4\ln(y)+a_{2.5}u^{5/2}+\sum_{i=1}^{8}a_iu^i}
{1+\sum_{i=1}^{4}b_iv^i}
\,.
\label{Axfit}
\ee
Setting $a_7=0$ in Eq.\ (\ref{Axfit}) eliminates the 
un-needed $y^{-2/3}$ dependence in the large-$y$ expansion. 
See Table \ref{tab:table6} for additional constraints on the parameters. 

Similarly, fitting of either the function $f$ or $\kappa$ requires 
incorporation of logarithmic terms in a modified  Pad\'e approximant.
We developed such a modified Pad\'e approximant as well (not shown here),
but fitting to it did not yield any material improvement with respect to the 
original Perrot fit \cite{Perrot.1979} 
which has very small relative errors (see Table \ref{tab:table1}).

The representation of  $\widetilde B_{\mathrm{x}}$ takes the form of 
a [20,20] Pad\'e approximant 
\be
\widetilde B_{\mathrm{x}}(y) = \frac{\sum_{i=2}^{10}a_iu^i}
{1+\sum_{i=1}^{10}b_iu^i}
\,,
\label{Bxfit}
\ee
where $a_1=a_9 = b_9=0$ to eliminate the $y^{2/3}$ and $y^{-2/3}$ terms
from the small-$y$ and large-$y$ expansions respectively. 
Additional constraints are imposed
on the parameters $a_2$, $b_8$ and $b_{10}$ to match the coefficients of 
the leading terms in 
Eqs.\ (\ref{small-y}) and (\ref{large-y}) (see Table \ref{tab:table7}).

The fitting function for $\eta_{1/2}$ is similar to that of  Eq.\ (\ref{Axfit})
due to the presence of the logarithmic term in the 
asymptotic expansion. Also, the 
constant $a_0$ and the ninth power 
of the variable $u$ were added in the numerator to satisfy the large-$y$
expansion Eq.\ (\ref{small-large-y-inv}), that is, 
\be
\eta_{1/2}(y) = \frac{a_{ln}\ln(y)+a_{2.5}u^{5/2}+\sum_{i=0}^{9}a_iu^i}
{1+\sum_{i=1}^{4}b_iv^i}
\,.
\label{etafit}
\ee
We take $a_8=0$ to eliminate a constant term in the large-$y$ expansion 
(see Eq.\ (\ref{small-large-y-inv}))
and $a_0=\ln(2/\sqrt{\pi})$, $a_{ln}=1$, $b_4=(2/3)^{2/3}a_9$
to match the expansion coefficients in Eq.\ (\ref{small-large-y-inv}).
Again, we found it beneficial to keep the $u^{5/2}$ term to reduce 
fitting errors.

Eventually, the fit to $\widetilde B$ has 12 independent parameters, 
while the $\widetilde C$, $\widetilde D$, $\widetilde E$ fits have 15   
and the fits to $\widetilde A_{\mathrm{x}}$ and 
$\widetilde B_{\mathrm{x}}$ have 10 and 14 parameters, 
respectively. The fitted inverse function $\eta_{1/2}$ has 
12 independent parameters.
For comparison, the Perrot fits \cite{Perrot.1979} have 15 and 
14 parameters for $\kappa$ and $\widetilde B$ respectively. 
The fits to $\widetilde B$ -- $\widetilde E$ 
from Ref.\ \cite{Geldart.Sommer.1985B} have, in order, 
10, 20, 9, and 14 parameters.
All parameters, including constrained ones, are listed in 
Tables \ref{tab:table2}-\ref{tab:table8}.  

\begin{table}
\caption{\label{tab:table1}
Results for present work (``PADE'') using LSF and
combined MARE fitting (see text) compared with fits from Perrot 
\cite{Perrot.1979}, Geldart and Sommer (GS) 
\cite{Geldart.Sommer.1985B},
Perrot-Dharma-wardana (PDW84) \cite{PDW84},
and Nilsson Eq.\ (8) in Ref.\ \cite{Nilsson.1973}
(same as Blakemore Eq.\ (38) in Ref.\ \cite{Blakemore1982}). 
Relative square error (RSE) (in \%) for the specified fitted function is
in column two.   
Mean absolute relative error (MARE) (in \%) for that 
fitted function ($F$) and for its first ($F'$) and 
second ($F''$) derivatives with respect to the variable $y$  
are in the third, fourth, and fifth columns. 
}
\begin{ruledtabular}
\begin{tabular}{llccc}
\tiny Fit & \tiny RSE($F$) & \tiny MARE($F$) & \tiny MARE($F'$) & \tiny MARE($F''$) \\
\hline
\underline{$\kappa(y)$}\\
Perrot  & 7.0$\cdot 10^{-10}$ & 0.0002 & 0.004 & 0.013 \\
\hline
\hline
\underline{$\widetilde B(y)$}\\
Perrot              & 2.4$\cdot 10^{-6}$ & 0.008 & 1.5 & 2.1 \\
GS\footnotemark[1]
                    & 5.4$\cdot 10^{-4}$ & 0.23    & 37 & 581 \\
Pad\'e (LSF)             & 1.0$\cdot 10^{-8}$ & 0.001 & 0.52 & 0.72 \\
Pad\'e (MARE fit)            & 4.5$\cdot 10^{-8}$ & 0.0004 & 0.017 & 0.055 \\
\hline
\hline
\underline{$\widetilde C(y)$}\\
GS\footnotemark[1]
                    & 1.5$\cdot 10^{-3}$ & 0.63   & 21   & 736 \\
Pad\'e (LSF)     & 2.3$\cdot 10^{-7}$ & 0.01 & 0.65 & 0.73 \\
Pad\'e (MARE fit)    & 2.5$\cdot 10^{-6}$ & 0.008  & 0.17  & 0.31\\
\hline
\hline
\underline{$\widetilde D(y)$}\\
GS\footnotemark[1]
                    & 1.2$\cdot 10^{-3}$ & 0.80 & 10 & 135 \\
Pad\'e (LSF)     & 8.8$\cdot 10^{-7}$ & 0.013 & 1.1 & 1.3 \\
Pad\'e (MARE fit)    & 3.3$\cdot 10^{-5}$ & 0.015 & 0.47 & 0.67 \\
\hline
\hline
\underline{$\widetilde E(y)$}\\
GS\footnotemark[1]
                    & 1.9$\cdot 10^{-3}$ & 10 & 6.7 & 47 \\
Pad\'e (LSF)     & 1.7$\cdot 10^{-6}$ & 0.037 & 1.1 & 1.4 \\
Pad\'e (MARE fit)    & 2.5$\cdot 10^{-5}$ & 0.027 & 0.50 & 0.93 \\
\hline
\hline
\underline{$\widetilde A_{\mathrm{x}}(y)$}\\
PDW84                & 5.9$\cdot 10^{-2}$ & 0.19  & 27 & 23 \\
Pad\'e (LSF)      & 3.3$\cdot 10^{-9}$ & 0.001 & 0.05 & 0.07 \\
Pad\'e (MARE fit)    & 1.7$\cdot 10^{-8}$ & 0.001 & 0.02 & 0.04 \\
\hline
\hline
\underline{$\widetilde B_{\mathrm{x}}(y)$}\\
Pad\'e (LSF)      & 1.3$\cdot 10^{-6}$ & 0.021 & 0.67 & 1.4 \\
Pad\'e (MARE fit)    & 1.0$\cdot 10^{-5}$ & 0.023 & 0.32 & 0.49 \\
\hline
\hline
\underline{$\eta_{1/2}(y)$}\\
Nilsson              & 0.7                 & 8.5     & 6.9     & 6.6 \\
Pad\'e (LSF)      & 1.3$\cdot 10^{-10}$ & 0.0015 & 0.0028 & 0.040 \\
Pad\'e (MARE fit)    & 4.0$\cdot 10^{-10}$ & 0.0009 & 0.0026 & 0.035 \\
\end{tabular}
\end{ruledtabular}
\footnotetext[1]{Fit from Ref.\ \cite{Geldart.Sommer.1985B}; errors calculated 
only  on the interval $y \in [0.02,25]$ used for fitting.}
\end{table}
%

The remaining issue is determination of the coefficients left undetermined 
after enforcement of the asymptotic criteria.
Numerical exploration led to the conclusion that ordinary LSF techniques
are not adequate for that task. The standard LSF criterion is to minimize the
squared error (SE) 
${\mathcal E}_{\mathrm {SE}} :=\sum_{i=1}^{N}( F_i - F_{\mathrm{fit},i})^2$,
which is related to the relative squared error (RSE)
\be
{\mathcal E}_{\mathrm {RSE}} :=  %
\frac{\sum_{i=1}^{N}( F_i - F_{\mathrm{fit},i})^2}{\sum_{i=1}^{N}F_i^2}  \; .
\label{RSE}
\ee 
However, a small SE (or a small RSE) obviously does not guarantee a small 
mean absolute relative error (MARE) for a function and the derivatives 
of that fitted function.  Both a more sensitive measure and one which
takes explicit account of at least the low-order derivatives is required. 
Hence we adopted fitting based on a weighted MARE criterion.
With the MARE for a function $F$ defined as 
\be
{\mathcal E}_{\mathrm {MARE}}[F] := N^{-1}\sum_{i=1}^{N} %
\frac{\vert F_i - F_{\mathrm{fit},i}\vert}{\vert F_i\vert} \; ,
\label{MARE}
\ee 
the weighted error we minimized was
\be
\omega [F] =10\times {\mathcal E}_{\mathrm{MARE}}[F]  %
+ {\mathcal E}_{\mathrm{MARE}}[F^\prime(y)] %
+ {\mathcal E}_{\mathrm{MARE}}[F^{\prime\prime}(y)] \; .
\label{wtdMARE}
\ee
Here $F$ is considered as a function of the variable $y$
and primes once again indicate the derivatives.  
The results do not depend sensitively on the relative
weight of the function MARE with respect to the derivative MAREs. 
The RSE and MARE results also depend rather weakly
on the choice of the $\Delta \eta$ for spacing the
reference data and on the size of the $\eta$ interval
(see previous sub-Section for the choice used here).

\subsection{Accuracy}

Table \ref{tab:table1} shows the RSE
for fitted functions and MARE for the functions and their
first and second derivatives for the current work with
pure LSF and with the weighted MARE optimization from
Eq.\ (\ref{wtdMARE}).  These are compared with    
the previous representations from Perrot \cite{Perrot.1979} 
and Geldart and Sommer \cite{Geldart.Sommer.1985B} fits. 
Perrot's fit for $\kappa(y)$ yields very small MARE values
for both the function and its first and second derivatives,
i.e., the fit provides an accurate analytical representation for 
$\kappa$ (or, equivalently $f$). 
Perrot's fit to ${\widetilde B}(y)$
yields a small MARE for the function, but the 
MAREs for the first and second derivatives are not small 
(1.5\% and 2.1\% correspondingly). The Geldart and Sommer fits 
provide acceptable MAREs for the representations of 
${\widetilde B}$, ${\widetilde C}$, and ${\widetilde D}$ , 
but not for $\widetilde E$, for which 
the MARE is 10\%. However, the GS fits fail completely for the derivatives,
with MARE values from 6.7\% up to 736\%.  

Our Pad\'e (and modified Pad\'e) fits based on the weighted MARE 
criterion provide generally superior results. MARE values for 
fitted functions are between 0.0004\% and 0.027\%.
MAREs for derivatives of $\widetilde B$ and $\widetilde A_{\mathrm{x}}$
are of the order of hundredths of percent.
For the derivatives of $\widetilde C$, $\widetilde D$, $\widetilde E$
and $\widetilde B_{\mathrm{x}}$ the  MAREs are 
less than one percent. 

\begin{figure}
\includegraphics*[width=7.5cm]{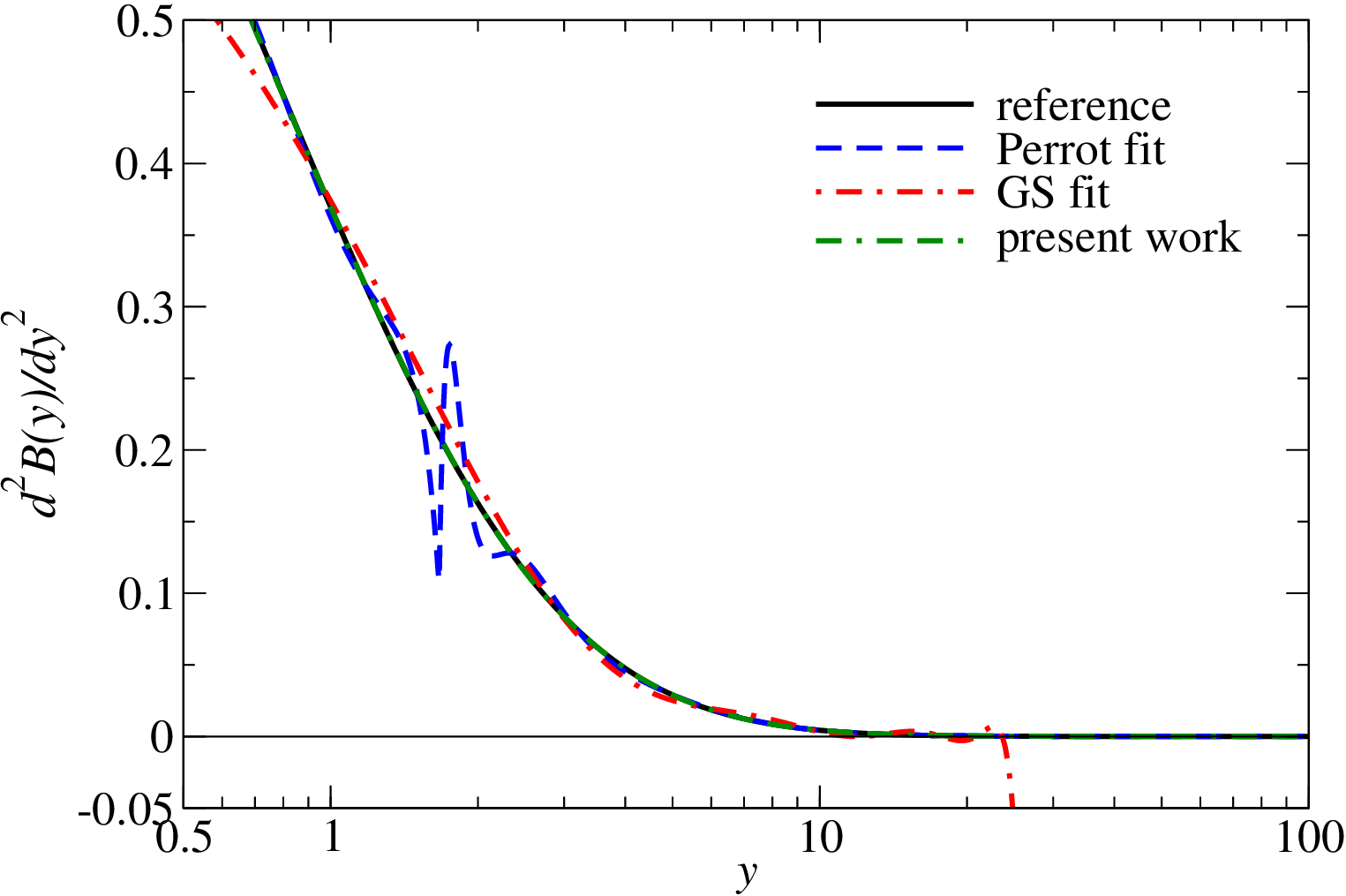}
\caption{
Second derivative $\widetilde B''(y)$ as a function of $y$. Comparison
between reference data and Perrot, GS, and weighted MARE Pad\'e fits.
}
\label{d2Bdy2}
\end{figure}

\begin{figure*}
\includegraphics*[width=7.5cm]{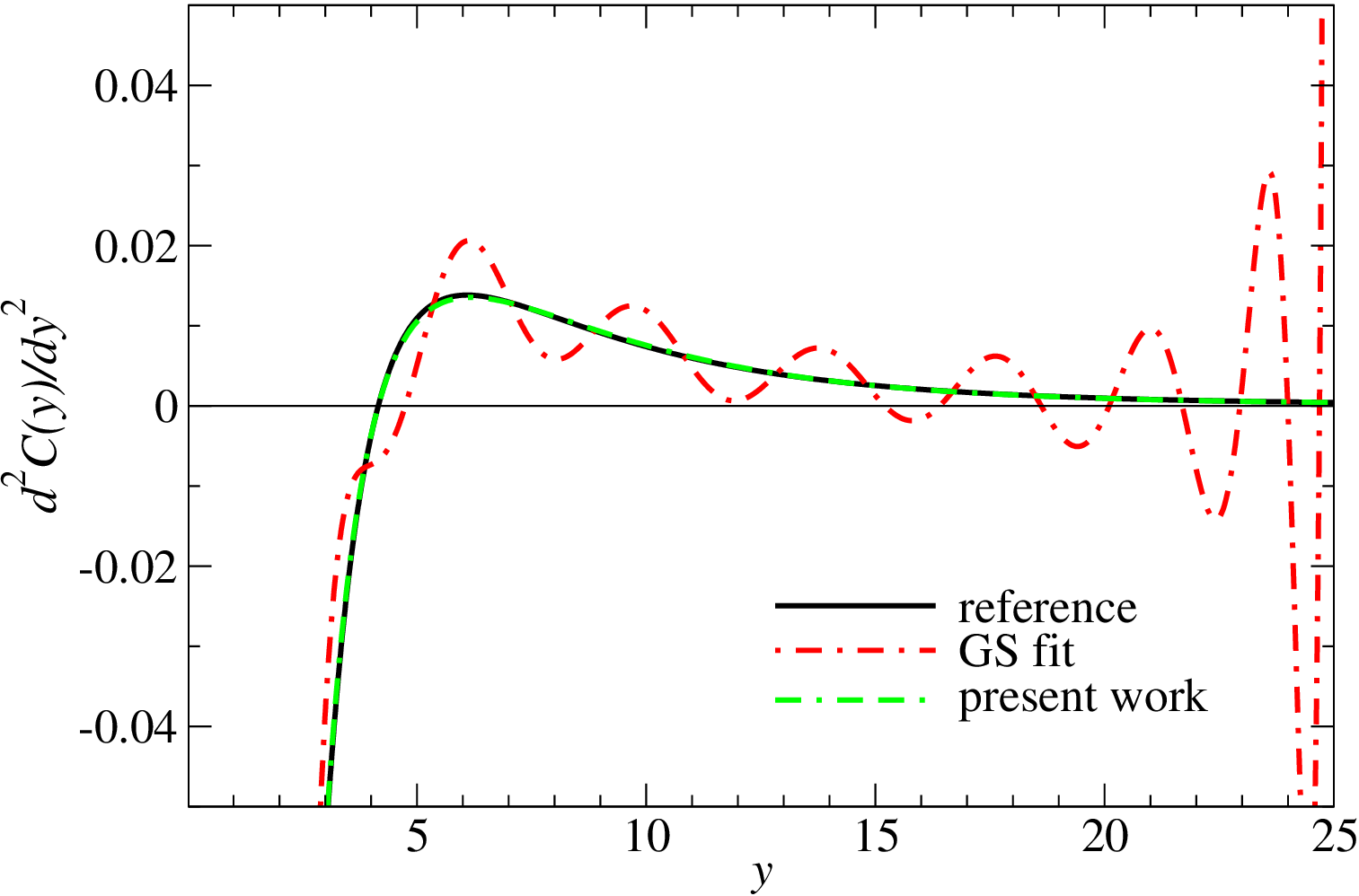}
\includegraphics*[width=7.5cm]{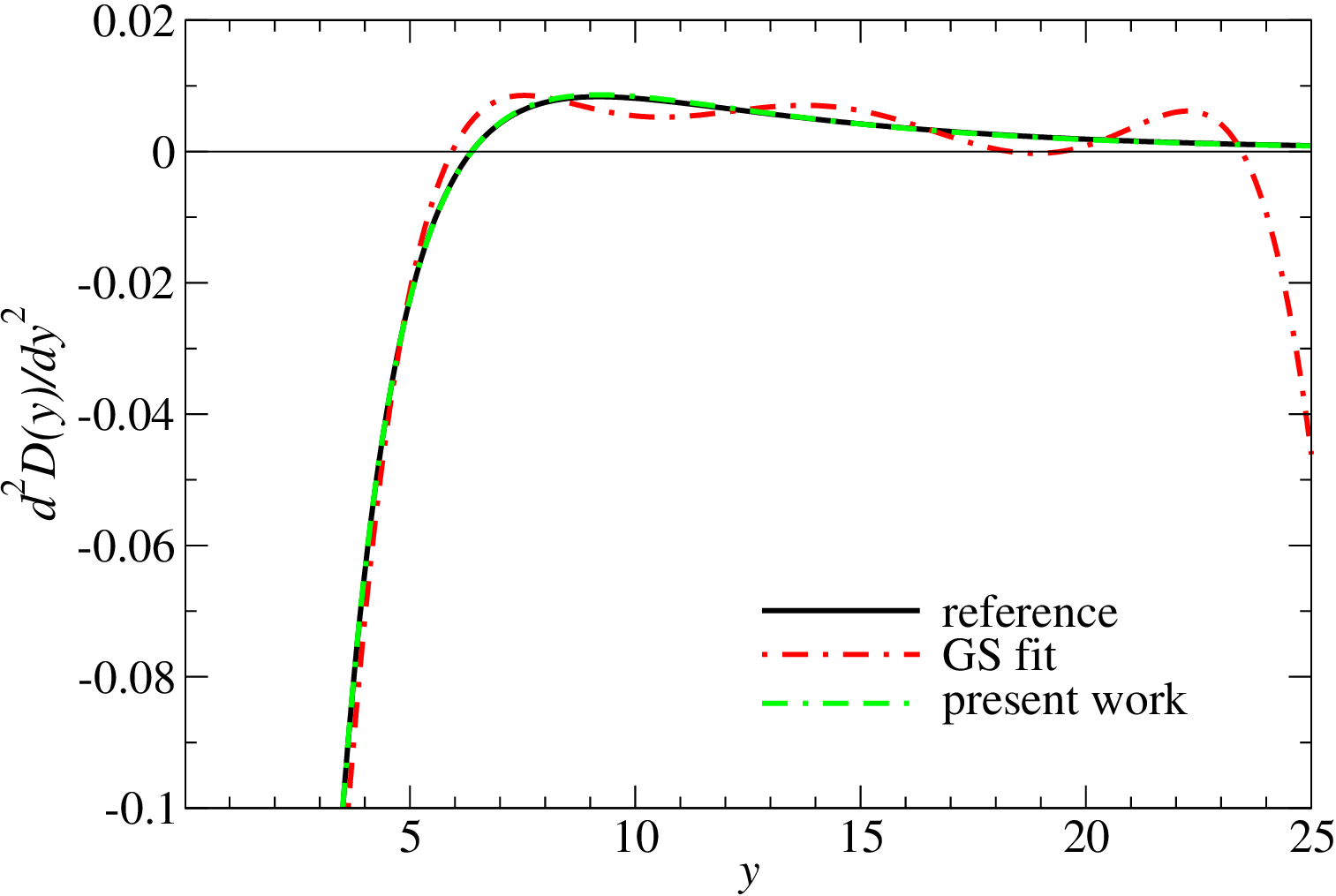}
\includegraphics*[width=7.5cm]{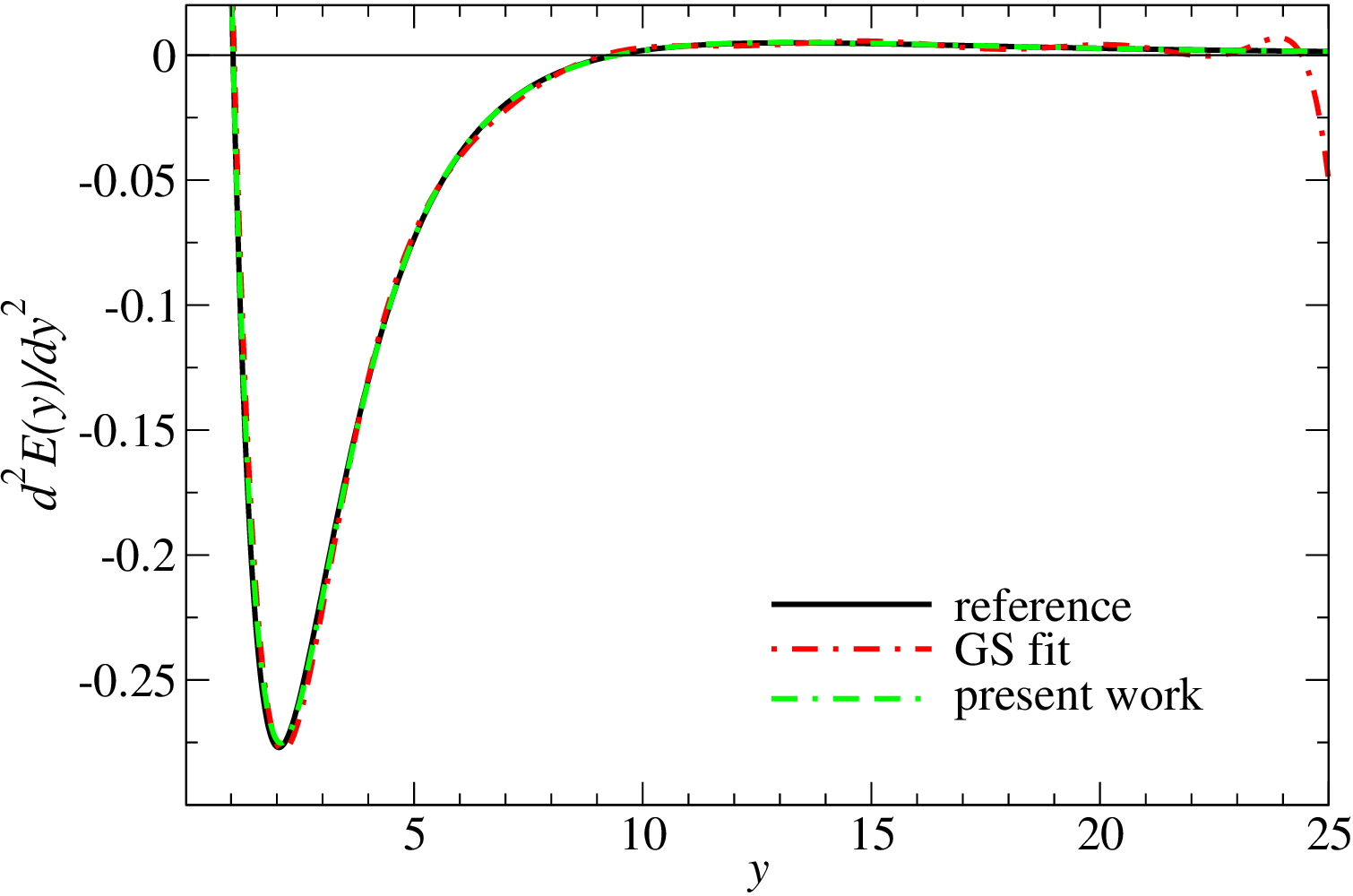}
\includegraphics*[width=7.5cm]{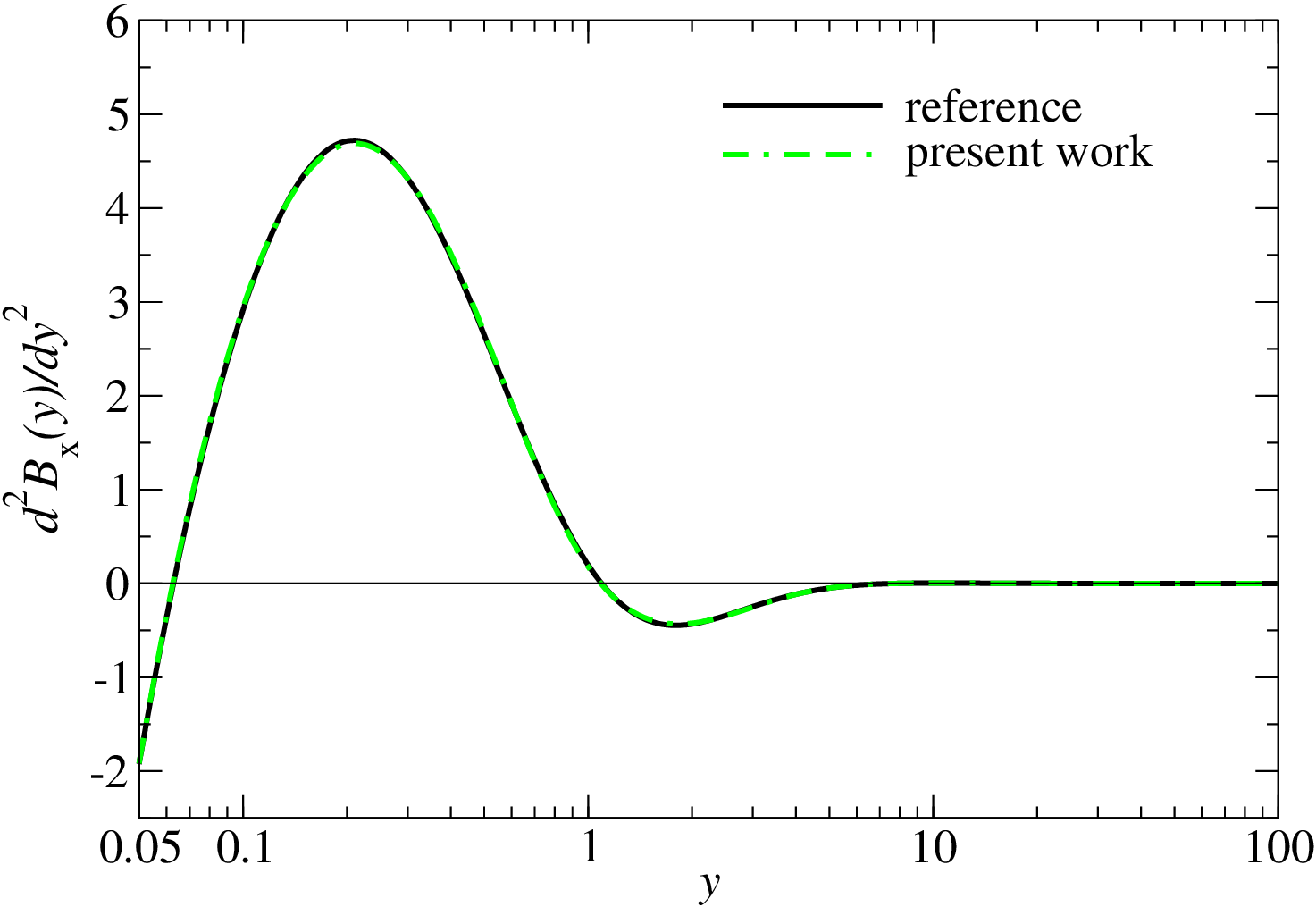}
\caption{
Second derivatives $\widetilde C''(y)$, $\widetilde D''(y)$,
$\widetilde E''(y)$ and $\widetilde B_{\mathrm{x}}''(y)$ as functions of $y$. 
Comparison between reference data and fits.
}
\label{d2CDEBxdy2}
\end{figure*}

Unsurprisingly, the second derivative is the most sensitive quantity 
to fitting errors. 
Fig.\ \ref{d2Bdy2} shows the second derivative $\widetilde B''(y)$
calculated from the Perrot, GS, and our weighted MARE Pad\'e fit 
compared to the reference calculated values.
Overall, the Perrot fit provides good agreement with the tabulated data 
except in the region near $y_0\approx 1.67$, 
where the second derivative has huge oscillations. 
The GS fit oscillates about the reference values over the 
entire range of $y$. In contrast, the weighted MARE Pad\'e fit 
is smooth and generally in excellent agreement 
with the tabulated data. Mean absolute relative errors 
for the second derivative 
$\widetilde B''(y)$ are 2.1\%, 580\% and 0.055\% for the Perrot, GS, and 
weighted MARE Pad\'e fits correspondingly.
Fig.\ \ref{d2CDEBxdy2} shows similar behavior from the GS fit for 
the second derivatives 
$\widetilde C''(y)$, $\widetilde D''(y)$ and $\widetilde E''(y)$.
Comparison between the 
calculated reference data and the weighted MARE Pad\'e fit 
for $\widetilde B_{\mathrm{x}}''(y)$ is excellent (see Fig. \ \ref{d2CDEBxdy2})
with an MARE of  0.55\%.

\begin{figure*}
\includegraphics*[width=7.5cm]{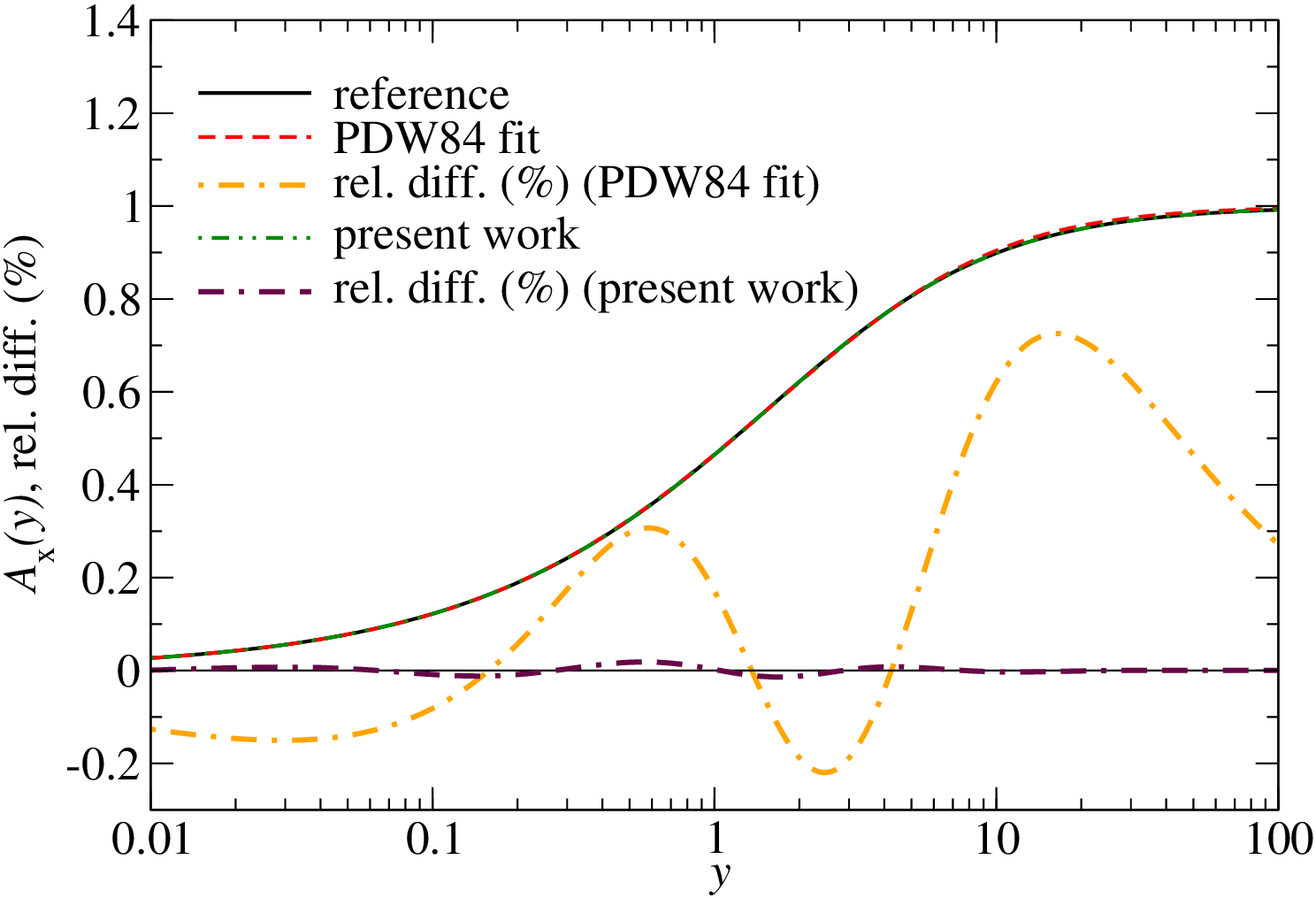}
\includegraphics*[width=7.5cm]{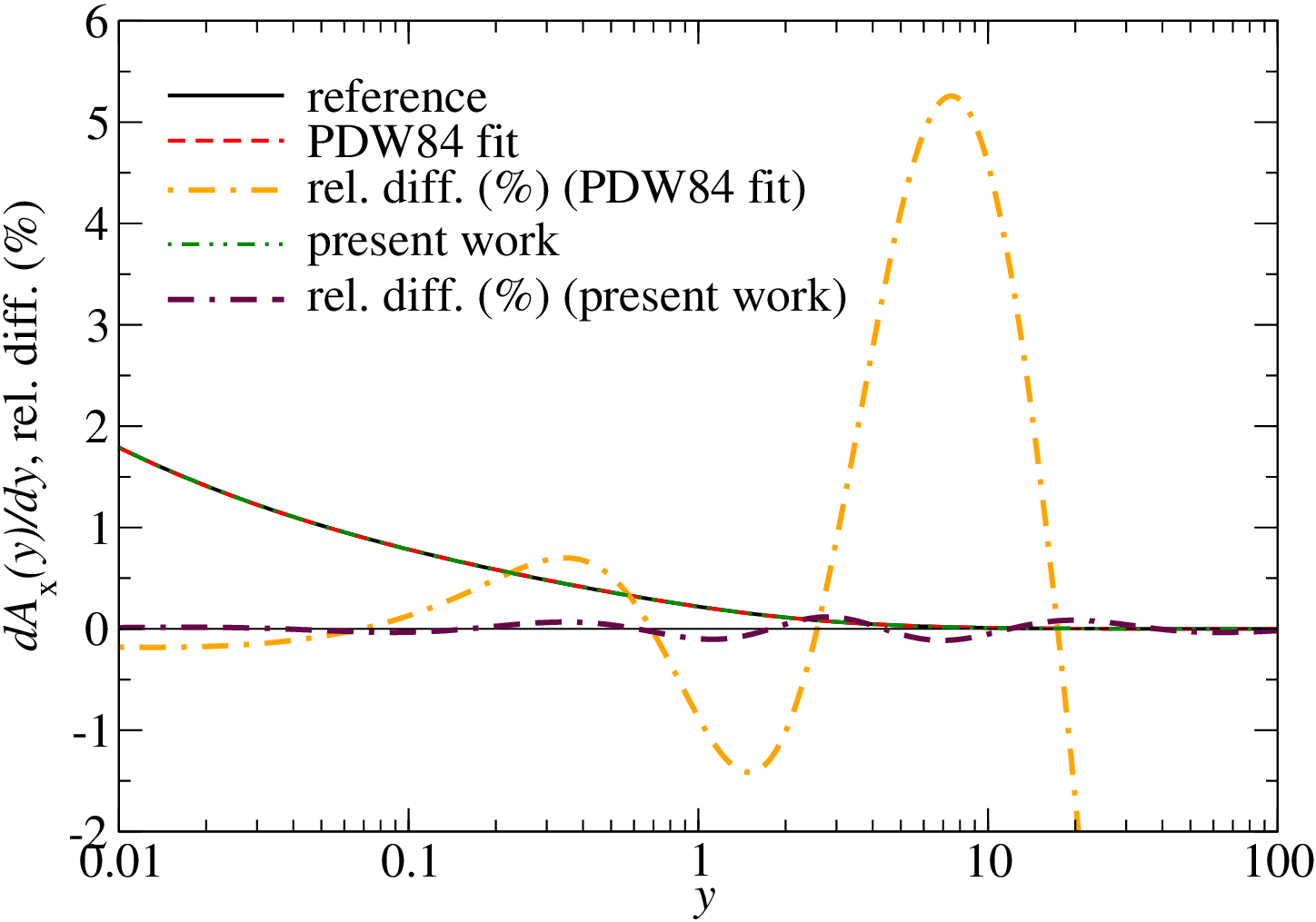}
\includegraphics*[width=7.5cm]{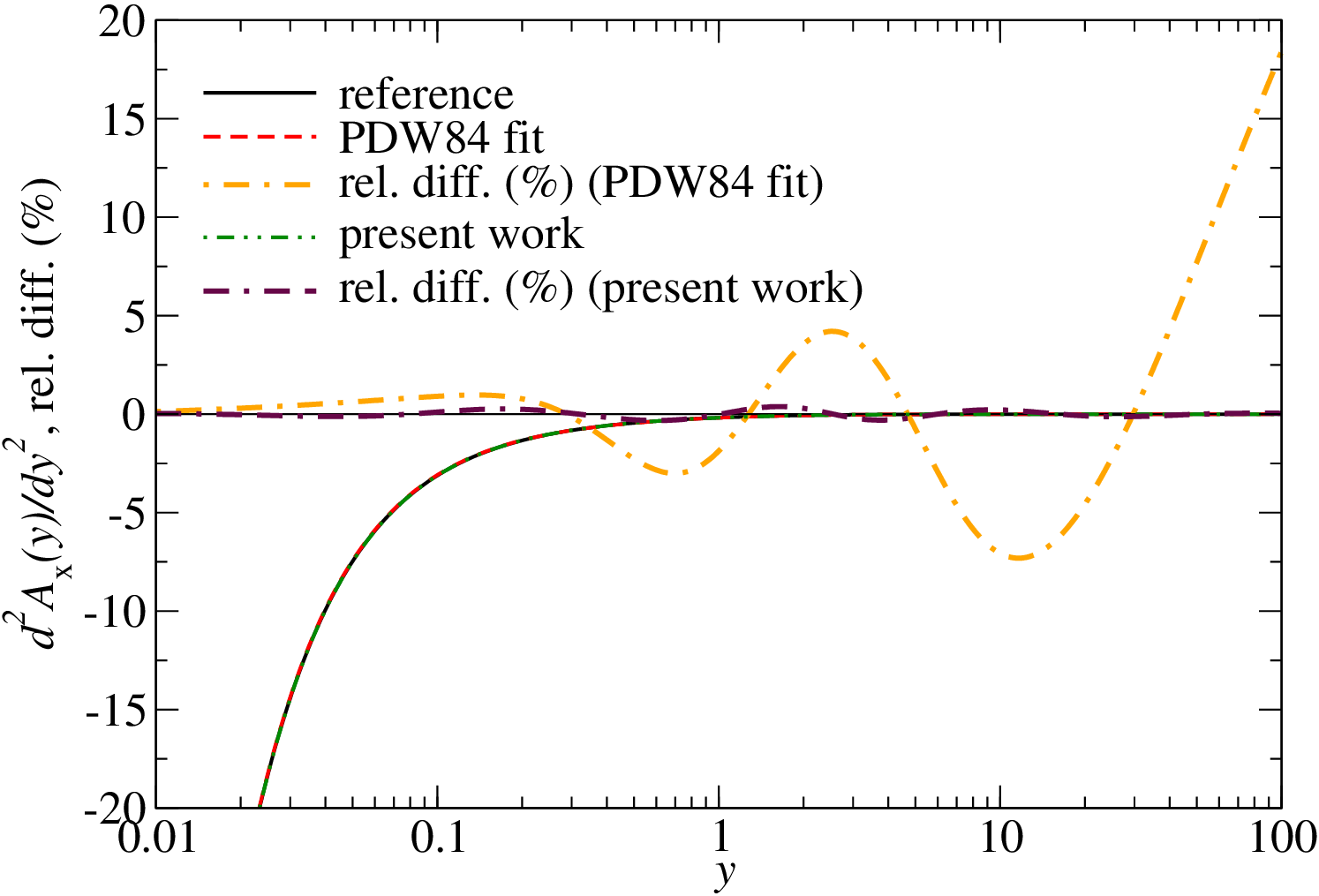}
\includegraphics*[width=7.5cm]{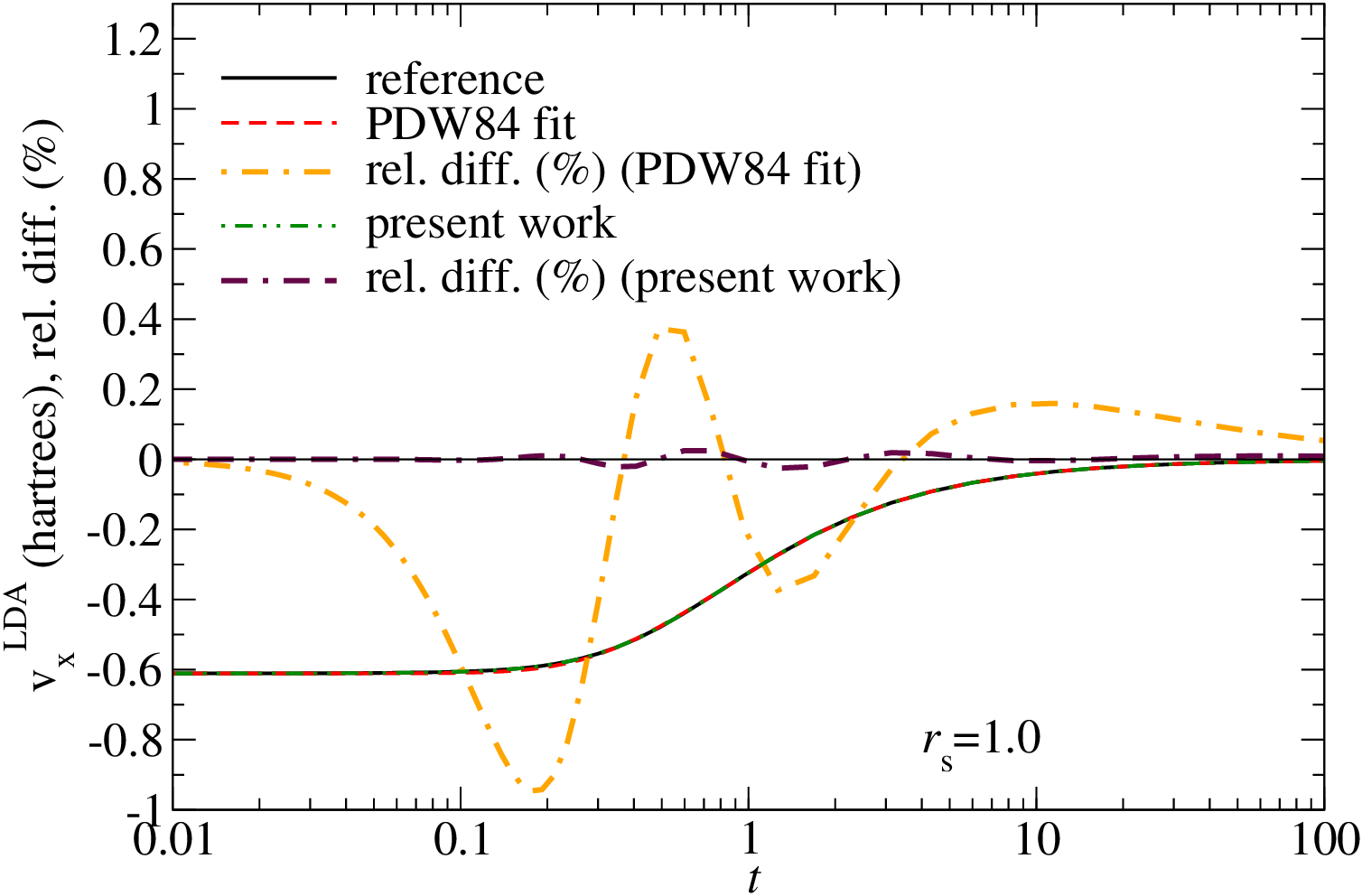}
\caption{
Comparison between reference data and fits for $\widetilde A_{\mathrm{x}}(y)$,
$\widetilde A_{\mathrm{x}}'(y)$ and $\widetilde A_{\mathrm{x}}''(y)$
as functions of $y$. Relative differences between corresponding fits and 
reference data also are shown.  The lower-right panel shows corresponding 
X potentials (as functions of $t$)
and relative differences between reference $v_{\mathrm{x}}^{\mathrm{LDA}}$
and values calculated from fits for $r_{\mathrm{s}}=1.0$ bohr.
}
\label{Ax-fig}
\end{figure*}

In a similar vein, Fig.\ \ref{Ax-fig} compares reference data,
the Perrot-Dharma-wardana (PDW84) \cite{PDW84}, and modified Pad\'e fits
for $\widetilde A_{\mathrm{x}}$ and its derivatives with respect to $y$.
The maximum relative error of the PDW84 fit to $\widetilde A_{\mathrm{x}}$ 
is about
0.7\% (upper-left panel) at $y\approx 20$ ($t\approx 0.1$). 
That seemingly small error might be not negligible for 
certain calculations.  As an example consider  use of 
a fit to $\widetilde A_{\mathrm{x}}$ 
for calculation of the LDA correlation free-energy per particle 
($f_{\mathrm{c}}^{\mathrm{LDA}}/n$)
from the LDA exchange-correlation free-energy,
$f_{\mathrm{c}}^{\mathrm{LDA}}/n\equiv
(f_{\mathrm{xc}}^{\mathrm{LDA}}-f_{\mathrm{x}}^{\mathrm{LDA}})/n$. 
At $t=0.125$ ($y=15.085$), $r_{\mathrm{s}}=1.0$ the PDW84 fit gives
$f_{\mathrm{c}}^{\mathrm{LDA}}/n=-0.5200+0.4309=-0.0891$ hartree 
(see Table S1 in Supplemental Material for Ref.\ \cite{LSDA-PIMC}). 
The accurate modified Pad\'e fit introduced here gives 
$f_{\mathrm{c}}^{\mathrm{LDA}}/n=-0.5200+0.4278=-0.0922$ hartree, 
that is, about 3\% lower.  
Moreover, despite the reasonably accurate fit provided by the PDW84 form
(MARE = 0.19\%), 
it does not provide accurate first and second derivatives.
Fig.\ \ref{Ax-fig}  also shows relative differences between 
reference data and corresponding fits for those derivatives.   
Errors in the first derivative of $\widetilde A_{\mathrm{x}}(y)$ propagate
as errors in the corresponding exchange potential
\be
v_{\mathrm{x}}^{\mathrm{LDA}}(n,t)=\frac{\partial e_{\rm x}^{\rm LDA}(n)}{\partial n} \widetilde A_{\mathrm{x}}(y(t))
+ e_{\rm x}^{\rm LDA}(n) \frac{\partial\widetilde A_{\mathrm{x}}(y(t))}{\partial t}\frac{\partial t}{\partial n}
\,.
\label{vx}
\ee
Consequences of this error propagation are shown in the lower-right
panel of Fig.\ \ref{Ax-fig}, which compares exchange potentials
from the present fit and the PDW84 fit with reference data, all  
calculated for $r_s=(3/4\pi n)^{1/3}=1$ bohr.
The relative errors for $v_{\mathrm{x}}^{\mathrm{LDA}}$ from the PDW84 fit are 
near 1\% for $0.1<t<2$.
Errors in $v_{\mathrm{x}}^{\mathrm{LDA}}$ are reduced as compared to 
corresponding errors in the first derivative of $\widetilde A_{\mathrm{x}}$ 
because the X potential is dominated by the first term in 
Eq.\ (\ref{vx}).
Results for other values of $r_s$ are similar.
The fit based on the modified Pad\'e approximant Eq.\ (\ref{Axfit}) 
provides practically perfect
agreement with the reference data for  
$\widetilde A_{\mathrm{x}}$, its first two derivatives, and for 
the corresponding exchange potential.


%

%

%

The last three lines of Table \ref{tab:table1} show
errors for fits to the inverse function $\eta_{1/2}$. 
The Nilsson fit \cite{Nilsson.1973,Blakemore1982} proves to be
only a  
semi-quantitative approximation at best, with MAREs 
close to 10\% for both the function and its derivatives. 
Our modified Pad\'e MARE fit provides very accurate results, with MARE 
for the function only 0.0009\%. The LSF provides practically 
the same accuracy.

\section{Conclusions}

We have developed accurate analytical representations for 
six combinations of Fermi-Dirac integrals 
and for one of the most important inverse functions arising in 
finite-T DFT.  The representations  have either the form of 
Pad\'e approximants or  modified Pad\'e approximants as needed to 
accommodate required asymptotic behaviors. Parameters in 
the analytical forms 
are constrained to reproduce correctly the leading terms in 
the asymptotic expansions for both extremes of the variable $y$. 

The new representations enable fast, accurate evaluation
of  the required FD integral combinations and their derivatives 
with improved accuracy compared to previously published
fits. MAREs for the new representations are
of the order of hundredths of a per cent in worse cases. 
The new representations furthermore are intrinsically continuous
with continuous derivatives of any order. 
The only previously published fit which has suitably small errors
for both the function and its derivatives is that by 
Perrot for $f=3\kappa/(5t)$. 
A set of {\sc Fortran} subroutines for all the new improved fits 
and corresponding derivatives is available
by download from http://www.qtp.ufl.edu/ofdft and by request to
the authors.

\section{Acknowledgments}
We thank J.W.\ Dufty for helpful comments.  We thank the University 
of Florida High-Performance Computing Center for 
computational resources and technical support. 
This work was supported by the U.S.\ Dept.\ of Energy TMS  
grant DE-SC0002139.

\appendix*

\begin{center} 
\textbf{Appendix}
\end{center}


The coefficients of the Pad\'e fits to $\widetilde B$, $\widetilde C$,  
$\widetilde D$, $\widetilde E$, $\widetilde A_{\mathrm{x}}$,
$\widetilde B_{\mathrm{x}}$ and $\eta_{1/2}$ as functions of the $y$
variable  
(see Eqs.\ (\ref{Bfit})-(\ref{etafit}))
are given in Table \ref{tab:table2}-\ref{tab:table8}.
The second column shows the constraints imposed on 
corresponding coefficients.  A ``yes'' signifies that 
the coefficient was constrained to be $0$, $1$, or $3$ as the case may be. 

\begin{table}[h]
\caption{\label{tab:table2}
Coefficients in fit to $\widetilde B(y)$.
}
\begin{ruledtabular}
\begin{tabular}{lrr}
coefficient & constraint & \rm value \\
\hline
$a_{ 0}$ &    yes                        &    3.0000000000000000 \\
$a_{ 1}$ &   $-3/\sqrt{2\pi}$            &   -1.1968268412042982 \\
$a_{ 2}$ &                               &  427.3949714847699966 \\
$a_{ 3}$ &                               & -170.1211444343163919 \\
$a_{ 4}$ &                               &   31.7020753506680002 \\
$a_{ 5}$ &                               &    3.3713851108273998 \\
$a_{ 6}$ &                               &    2.2529104734200001 \\
$a_{ 7}$ &  yes                          &    0.0000000000000000 \\
$a_{ 8}$ &                               &    0.0202417083225910 \\
\hline
$b_{ 1}$ &  yes                          &    0.0000000000000000 \\
$b_{ 2}$ &  yes                          &    0.0000000000000000 \\
$b_{ 3}$ &                               &  142.2807110810987865 \\
$b_{ 4}$ &  yes                          &    0.0000000000000000 \\
$b_{ 5}$ &                               &    0.5924932349226000 \\
$b_{ 6}$ &                               &  -18.0196644249469990 \\
$b_{ 7}$ &                               &    7.2322601129560002 \\
$b_{ 8}$ &                               &    0.1910870984626600 \\
$b_{ 9}$ &                               &    2.2522978973395000 \\
$b_{10}$ & $-2^{4/3}3^{-7/3}\pi^2b_{12}$ &   -0.0387826345397392 \\
$b_{11}$ &   yes                         &    0.0000000000000000 \\
$b_{12}$ &    $a_8$                      &    0.0202417083225910 \\
\end{tabular}
\end{ruledtabular}
\end{table}

\begin{table}[!t]
\caption{\label{tab:table3}
Coefficients in fit to $\widetilde C(y)$.
}
\begin{ruledtabular}
\begin{tabular}{lrr}
coefficient & constraint & \rm value \\
\hline
$a_{2.5}$&                   &   5.9265262369781002 \\
$a_{ 1}$ & $3^{5/3}2^{-5/3}$ &   1.9655560456566725 \\
$a_{ 2}$ &                   &  -0.5768378962095700 \\
$a_{ 3}$ &                   &  35.9130119576930014 \\
$a_{ 4}$ &                   &  41.1168867899709980 \\
$a_{ 5}$ &                   & -40.3677476700629967 \\
$a_{ 6}$ &                   &  59.6804384544149968 \\
$a_{ 7}$ &                   &  -0.3211461169282900 \\
$a_{ 8}$ &                   &   4.2815226867198000 \\
$a_{ 9}$ &                   &   0.0030385200207883 \\
$a_{10}$ &                   &   0.1596522984577500 \\
$a_{11}$ &  yes              &   0.0000000000000000 \\
$a_{12}$ &                   &   0.0056843727998872 \\
\hline				  
$b_{ 1}$ &                   & 26.5710993646139997 \\
$b_{ 2}$ &                   & 20.1172145257690005 \\
$b_{ 3}$ &                   & 17.6858602829550016 \\
$b_{ 4}$ &                   &  3.6467884940180002 \\
$b_{ 5}$ & $(a_{10}-C\times b_6)\times b_6/a_{12}$\tablenotemark[1]
                             &  0.1365086602125932 \\
$b_{ 6}$ &  $a_{12}$         &  0.0056843727998872 \\
\end{tabular}			  
\end{ruledtabular}
\tablenotetext[1]{$C=17\times 2^{-5/3}3^{-7/3}\pi^2$.}
\end{table}

\begin{table}
\caption{\label{tab:table4}
Coefficients in fit to $\widetilde D$(y).
}
\begin{ruledtabular}
\begin{tabular}{lrr}
coefficient & constraint & \rm value \\
\hline
$a_{2.5}$&                   &    0.4524584047298800\\
$a_{ 1}$ & $2^{1/3}3^{-1/3}$ &    0.8735804647362989\\
$a_{ 2}$ &                   &    0.0300776040166210\\
$a_{ 3}$ &                   &   14.3828916532949993\\
$a_{ 4}$ &                   &    1.8670041583370001\\
$a_{ 5}$ &                   &   37.9149736744980004\\
$a_{ 6}$ &                   &    0.7589550686574100\\
$a_{ 7}$ &                   &   16.9530731446740006\\
$a_{ 8}$ &                   &    3.1373656916102002\\
$a_{ 9}$ &                   &    0.0241382844920020\\
$a_{10}$ &                   &    0.1538471708464500\\
$a_{11}$ &  yes              &    0.0000000000000000\\
$a_{12}$ &                   &    0.0049093483855146\\
\hline			
$b_{ 1}$ &                   &   15.9188442750290005\\
$b_{ 2}$ &                   &   29.1916070884210015\\
$b_{ 3}$ &                   &   14.7377409947669999\\
$b_{ 4}$ &                   &    3.1005334835656000\\
$b_{ 5}$ & $(a_{10}-C\times b_6)\times b_6/a_{12}$\tablenotemark[1]
                             &    0.1178771827774314\\
$b_{ 6}$ &  $a_{12}$         &    0.0049093483855146\\
\end{tabular}			  
\end{ruledtabular}
\tablenotetext[1]{$C=413\times 2^{-2/3}3^{-16/3}\pi^2$.}
\end{table}

\begin{table}
\caption{\label{tab:table5}
Coefficients in fit to $\widetilde E(y)$.
}
\begin{ruledtabular}
\begin{tabular}{lrr}
coefficient & constraint & \rm value \\
\hline
$a_{2.5}$& $3^{8/3}2^{-25/6}\pi^{-1/2}$&   0.5881075583333214\\
$a_{ 1}$ & yes                         &                  0.0\\
$a_{ 2}$ & yes                         &                  0.0\\
$a_{ 3}$ &                   &  -0.0132237512072000\\
$a_{ 4}$ &                   &   0.5865252375234600\\
$a_{ 5}$ &                   &   1.1120705517211000\\
$a_{ 6}$ &                   &   2.2626091489173001\\
$a_{ 7}$ &                   &   2.6723837550020000\\
$a_{ 8}$ &                   &   0.3385116347002500\\
$a_{ 9}$ &                   &   0.0038743130529412\\
$a_{10}$ &                   &   0.0108166294882730\\
$a_{11}$ &  yes              &   0.0000000000000000\\
$a_{12}$ &                   &   0.0003699371553596\\
\hline			
$b_{ 1}$ &                   &   2.8191769574094998\\
$b_{ 2}$ &                   &   7.4555425143053000\\
$b_{ 3}$ &                   &   2.5142144377484001\\
$b_{ 4}$ &                   &   0.3944764252937600\\
$b_{ 5}$ & $(a_{10}-C\times b_6)\times b_6/a_{12}$\tablenotemark[1]
                             &   0.0066524832876068\\
$b_{ 6}$ &  $a_{12}$         &   0.0003699371553596\\
\end{tabular}			  
\end{ruledtabular}
\tablenotetext[1]{$C=47\times 2^{-5/3}3^{-7/3}\pi^2$.}
\end{table}

\begin{table}
\caption{\label{tab:table6}
Coefficients in fit to $\widetilde A_{\mathrm{x}}(y)$.
}
\begin{ruledtabular}
\begin{tabular}{lrr}
coefficient & constraint & \rm value \\
\hline
$a_{ln}$ &  $C\times b_4$\tablenotemark[1]                 
                             & -0.0475410604245741 \\
$a_{2.5}$&                   & -0.1065378473507800 \\
$a_{ 1}$ & $2^{4/3}3^{-4/3}$ &  0.5823869764908659 \\
$a_{ 2}$ &                   & -0.0068339509356661 \\
$a_{ 3}$ &                   & 11.5469239288490009 \\
$a_{ 4}$ &                   & -0.8465428870889800 \\
$a_{ 5}$ &                   & -0.1212525366470300 \\
$a_{ 6}$ &                   &  1.9902818786101000 \\
$a_{ 7}$ &  yes              &  0.0000000000000000 \\
$a_{ 8}$ &                   &  0.0744389046707120 \\
\hline				  
$b_{ 1}$ &                   & 19.9256144707979992 \\
$b_{ 2}$ &                   &  5.1663994545590004 \\
$b_{ 3}$ &                   &  2.0463164858237000 \\
$b_{ 4}$ & $a_{8}$           &  0.0744389046707120 \\
\end{tabular}			  
\end{ruledtabular}
\tablenotetext[1]{$C=-2^{4/3}3^{-10/3}\pi^2$.}
\end{table}

\begin{table}
\caption{\label{tab:table7}
Coefficients in fit to $\widetilde B_{\mathrm{x}}(y)$.
}
\begin{ruledtabular}
\begin{tabular}{lrr}
coefficient & constraint & \rm value \\
\hline
$a_{ 2}$ & $-3^{4/3}2^{-1/3}\pi^{-1/2}$ &-3.4341427276599950 \\
$a_{ 3}$ &   		                &-0.9066069544311700 \\
$a_{ 4}$ &   		                & 2.2386316137237001 \\
$a_{ 5}$ &   		                & 2.4232553178542000 \\
$a_{ 6}$ &   		                &-0.1339278564306200 \\
$a_{ 7}$ &   		                & 0.4392739633708200 \\
$a_{ 8}$ &   		                &-0.0497109675177910 \\
$a_{ 9}$ & yes  		        & 0.0000000000000000 \\
$a_{10}$ &   		                & 0.0028609701106953 \\
\hline	     					    
$b_{ 1}$ &   		                & 0.7098198258073800 \\
$b_{ 2}$ &   		                & 4.6311326377185997 \\
$b_{ 3}$ &   		                &-2.9243190977647000 \\
$b_{ 4}$ &   		                & 6.1688157841895004 \\
$b_{ 5}$ &   		                &-1.3435764191535999 \\
$b_{ 6}$ &   		                & 0.1576046383295400 \\
$b_{ 7}$ &   		                & 0.4365792821186800 \\
$b_{ 8}$ & $(a_{8}-C\times b_{10})\times b_{10}/a_{10}$\tablenotemark[1]
                                        &-0.0620444574606262 \\
$b_{ 9}$ & yes  		        & 0.0000000000000000 \\
$b_{10}$ & $a_{10}$  		        & 0.0028609701106953 \\
\end{tabular}			  
\end{ruledtabular}
\tablenotetext[1]{$C=2^{-2/3}3^{-1/3}\pi^2$.}
\end{table}

\begin{table}
\caption{\label{tab:table8}
Coefficients in fit to $\eta_{1/2}(y)$.
}
\begin{ruledtabular}
\begin{tabular}{lrr}
coefficient & constraint & \rm value \\
\hline
$a_{ln}$ &           yes     &  1.0                \\
$a_{2.5}$&                   &  -1.2582793945794000\\
$a_{ 0}$ &$\ln(2/\sqrt{\pi})$&   0.1207822376352453\\
$a_{ 1}$ &                   &   0.0233056178489510\\
$a_{ 2}$ &                   &   1.0911595094936000\\
$a_{ 3}$ &                   &  -0.2993063964300200\\
$a_{ 4}$ &                   &  -0.0028618659615192\\
$a_{ 5}$ &                   &   0.5051953653801600\\
$a_{ 6}$ &                   &   0.0419579806591870\\
$a_{ 7}$ &                   &   1.3695261714367000\\
$a_{ 8}$ &  yes              &   0.0000000000000000\\
$a_{ 9}$ &                   &   0.2685157355131100\\
\hline				  
$b_{ 1}$ &                   &   0.0813113962506270\\
$b_{ 2}$ &                   &   1.1903358203098999\\
$b_{ 3}$ &                   &   1.1445576113258000\\
$b_{ 4}$ & $(2/3)^{2/3}a_9$  &   0.2049158578610270\\
\end{tabular}			  
\end{ruledtabular}
\end{table}



\end{document}